%% file: main.tex
\documentclass[10pt,conference,letterpaper]{IEEEtran}

\usepackage{cite}

\usepackage{amsmath,amssymb}
\usepackage{graphicx}
\usepackage{textcomp}
\usepackage{xcolor}
\usepackage{pifont}
\usepackage{booktabs}
\usepackage{makecell}
\usepackage{threeparttable}
\usepackage{adjustbox}
\usepackage[caption=false,font=footnotesize]{subfig}
\usepackage{listings}
\usepackage{wasysym}
\usepackage{enumitem}
\usepackage{xspace}
\usepackage[hidelinks]{hyperref}

\AtBeginDocument{}

\hypersetup{
  pdftitle={DeepStack: Facilitating Co-Design Exploration of 3D DRAM-Stacked Accelerators for Distributed LLM Inference},
  pdfauthor={Zhiwen Mo; Guoyu Li; Hao (Mark) Chen; Yu Cheng; Zhengju Tang; Qianzhou Wang; Lei Wang; Shuang Liang; Lingxiao Ma; Yuxiao Guo; Wayne Luk; Jilong Xue; Hongxiang Fan},
  pdfkeywords={3D DRAM-stacked accelerators, design space exploration, distributed large language model inference, performance modeling}}

\newcommand{\oursys}{{DeepStack}\xspace}
\newcommand{\secref}[1]{\S~\ref{#1}}
\newcommand{\figref}[1]{Fig.~\ref{#1}}
\newcommand{\tabref}[1]{Table~\ref{#1}}
\newcommand{\code}[1]{\texttt{#1}}

\newif\ifarxivversion
\arxivversiontrue

\ifarxivversion
\makeatletter
\def\ps@IEEEtitlepagestyle{%
  \def\@oddhead{\hfil\raisebox{-8pt}[0pt][0pt]{\normalfont\fontsize{9}{10}\selectfont\itshape
    Accepted at the 59th IEEE/ACM International Symposium on Microarchitecture (MICRO 2026)}\hfil}%
  \let\@evenhead\@empty
  \let\@oddfoot\@empty
  \let\@evenfoot\@empty}
\makeatother
\input{tex/arxiv_badges}
\fi

\ifarxivversion
\title{{\huge DeepStack: Facilitating Co-Design Exploration of 3D\\
DRAM-Stacked Accelerators for Distributed LLM Inference}}
\else
\title{\huge DeepStack: Facilitating Co-Design Exploration of 3D DRAM-Stacked Accelerators for Distributed LLM Inference}
\fi

\author{%
\IEEEauthorblockN{%
Zhiwen Mo\textsuperscript{1},
Guoyu Li\textsuperscript{1},
Hao (Mark) Chen\textsuperscript{1},
Yu Cheng\textsuperscript{2},
Zhengju Tang\textsuperscript{2},
Qianzhou Wang\textsuperscript{1},
Lei Wang\textsuperscript{2},\\
Shuang Liang\textsuperscript{1},
Lingxiao Ma\textsuperscript{3},
Yuxiao Guo\textsuperscript{3},
Wayne Luk\textsuperscript{1},
Jilong Xue\textsuperscript{3}, and
Hongxiang Fan\textsuperscript{1}}
\IEEEauthorblockA{%
\textsuperscript{1}Imperial College London\quad
\textsuperscript{2}Peking University\quad
\textsuperscript{3}Tile-AI}}

\begin{document}

\maketitle
\input{tex/0_abstract_v3}

\input{tex/1_intro_v3}

\input{tex/2_background_v4}
\input{tex/3_design_v3}

\input{tex/4_evaluation_v3}

\input{tex/5_case_study_v4}
\input{tex/7_related_v3}
\input{tex/8_conclusion_v4}

\section*{Acknowledgment}
This work is supported by ARIA through its
\href{https://aria.org.uk/opportunity-spaces/nature-computes-better/scaling-compute/}{Scaling Compute programme}.
The support of the UK EPSRC (Grants EP/V028251/1, EP/S030069/1, and
EP/X036006/1), UKRI (Grant 256), KIAT, and Broadcom is gratefully acknowledged.

\clearpage
\input{tex/artifact_appendix_camera_ready}

\clearpage
\IEEEtriggeratref{127}
\bibliographystyle{IEEEtran}
\bibliography{refs}

\end{document}

%% file: tex/arxiv_badges.tex
\RequirePackage{eso-pic}
\AddToShipoutPictureFG*{%
  \AtPageUpperLeft{%
    \put(\LenToUnit{\dimexpr1in+\hoffset+\oddsidemargin+\textwidth+13mm\relax},%
         \LenToUnit{-16.8mm}){%
      \makebox[0pt][r]{%
        \href{https://www.microarch.org/micro59/submit/artifacts.php}{%
          \includegraphics[height=12.5mm]{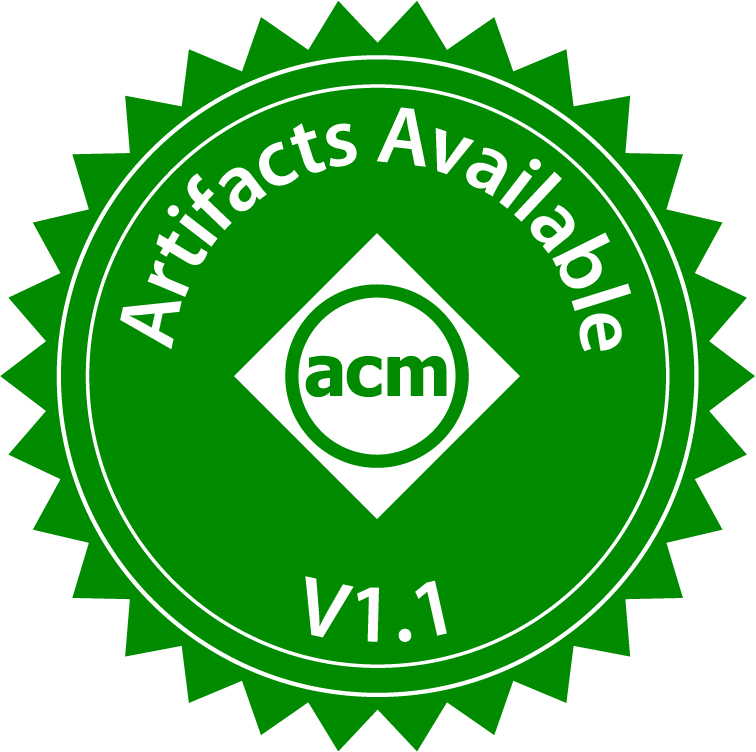}}%
        \hspace{1mm}%
        \href{https://www.microarch.org/micro59/submit/artifacts.php}{%
          \includegraphics[height=12.5mm]{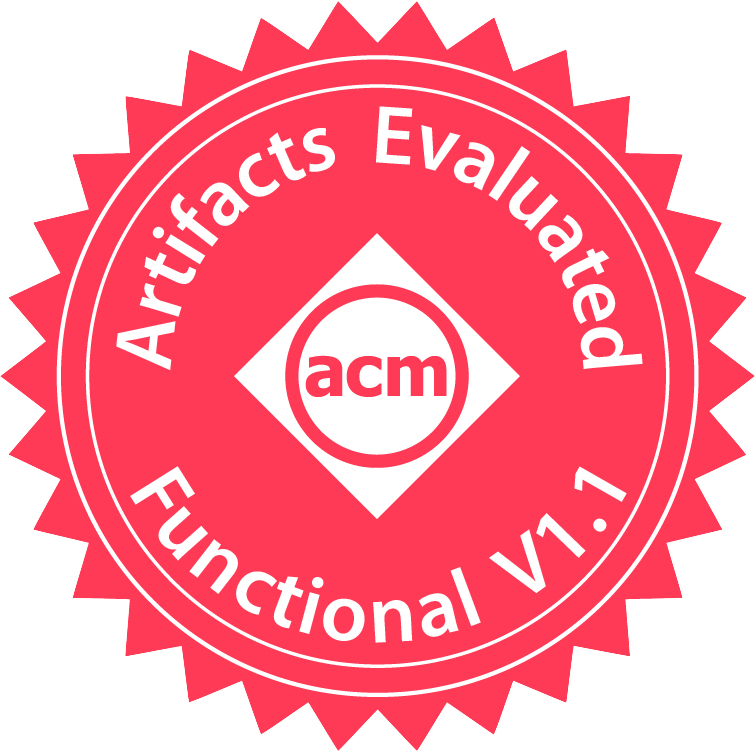}}%
        \hspace{1mm}%
        \href{https://www.microarch.org/micro59/submit/artifacts.php}{%
          \includegraphics[height=12.5mm]{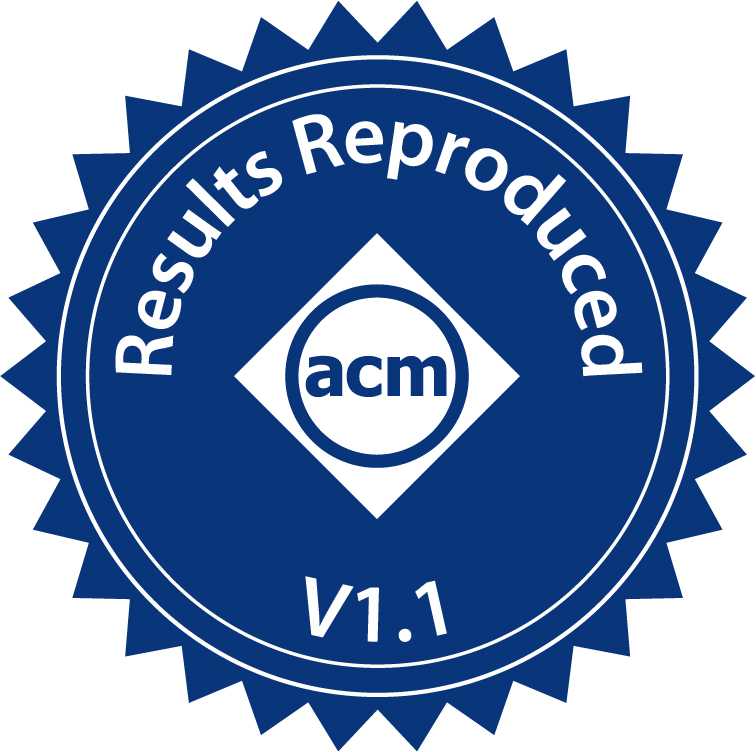}}%
      }%
    }%
  }%
}

%% file: tex/0_abstract_v3.tex
\begin{abstract}
Advances in hybrid bonding and packaging have driven growing interest in 3D DRAM-stacked AI accelerators.
As LLMs scale to hundreds of billions or trillions of parameters, distributed inference across multiple 3D chips has become essential for future AI serving.
This trend makes cross-stack co-design critical: system-level parallelization and scheduling choices are tightly coupled with low-level hardware characteristics such as memory organization, interconnect, and thermal constraints.
To enable early-stage co-design of distributed 3D AI systems, we present \oursys{}, an accurate performance model and efficient design space exploration (DSE) framework for distributed 3D-stacked LLM inference.
At the hardware level, \oursys{} captures fine-grained 3D memory characteristics, such as transaction-aware bandwidth, bank activation constraints, buffering limitations, and thermal–power modeling. At the system level, \oursys{} incorporates comprehensive parallelization strategies and execution scheduling for distributed LLM inference.
With novel modeling techniques such as dual-stage network abstraction and tile-level compute–communication overlap, we achieve up to $100{,}000\times$ faster runtime over state-of-the-art simulators at comparable accuracy. We cross-validate the modeling accuracy against our in-house 3D designs, NS-3 backend (2.12\% error), and vLLM serving on 8$\times$B200 GPUs (12.92\% error).
Combining with hierarchical design space search, \oursys{} enables efficient exploration over ${\sim}2.5 \times 10^{14}$ design points spanning across the number of 3D-stacked DRAM layers, DRAM vertical connectivity, interconnect, compute-memory allocation, and distributed scheduling under thermal and area constraints.
A search-space ablation shows that restricted DSE baselines can miss up to \mbox{$9.5\times$} modeled throughput.
Beyond modeling and DSE, we demonstrate that \oursys{} can be leveraged to derive design implications for distributed 3D AI systems, guiding performance optimization across the stack.
Source code and artifacts are available at
\href{https://github.com/tile-ai/DeepStack/tree/ae}{%
  \nolinkurl{https://github.com/tile-ai/DeepStack/tree/ae}}
\end{abstract}

%% file: tex/1_intro_v3.tex
\section{Introduction}

The scaling laws of LLMs~\cite{kaplan2020scaling,hoffmann2022training} have pushed model sizes beyond a trillion parameters~\cite{qwen3max,deepseekai2026deepseekv4,kimi2026kimik3}.
This growth, combined with agentic tasks and increasingly long context windows~\cite{liu2025comprehensive}, drives memory demand and bandwidth requirements beyond the capabilities of a single conventional accelerator, especially under batch-serving workloads in cloud deployments.
As a result, \textbf{distributed LLM inference} across multiple accelerators and nodes has become a fundamental requirement~\cite{Patel2025InferenceMAX,kwon2023vllm,zheng2024sglang,reddi2020mlperf}.
Emerging trends such as test-time compute scaling~\cite{snell2024scaling,muennighoff2025s1,liu2025can, chen2025rethinking, chen2026fasttts} and long-horizon agentic workloads~\cite{zhang2025survey,deng2025swepro,du2026deepresearch} can further intensify pressure on memory subsystems.

Recent breakthroughs in advanced packaging, such as high-density hybrid bonding~\cite{hu2021development,zhou2024research}, backside power delivery~\cite{hafez2023intel,ryckaert2019extending,jourdain2022buried}, and localized thermal management~\cite{salvi2021review,ao2024through-chip-cooling}, have driven active research towards 3D DRAM-stacked accelerator designs.
By vertically integrating DRAM with compute dies using hybrid bonding and fine-pitch through-silicon vias (TSVs)~\cite{sharda2025system}, these architectures provide significantly higher bandwidth and near-processor memory capacity, fundamentally mitigating memory bottlenecks versus traditional 2.5D designs.
This has also drawn increasing industry attention for large-scale distributed 3D-stacked AI systems~\cite{zhao2025insights-into-deepseek-v3,Ma2026ChallengesAR}.
However, the compounded complexity at both the system and hardware levels (up to $2.5 \times 10^{14}$ configurations) has hindered the development of accurate and efficient DSE frameworks for \emph{distributed} 3D-stacked AI systems, despite their importance in guiding early-stage design decisions.
Unlike modeling a single 3D chip, building an accurate and efficient DSE framework for multi-chip distributed 3D systems introduces several unique challenges:

\begin{itemize}[leftmargin=*]
\item \textit{\textbf{Chip-level accurate modeling with unique characteristics of 3D-stacked hardware.}}
3D-stacked DRAM provides significantly higher bandwidth, but fully utilizing it demands proportionally larger on-chip buffering (\textbf{Little's Law}~\cite{nvidia2025gtc_s72683}) and careful handling of bank-level access semantics such as large transaction sizes and high concurrency requirements~(\secref{subsubsec:challenge_dram_transaction_size}).
Vertical stacking also exacerbates thermal dissipation challenges, introducing power-constrained regimes that must be co-modeled with performance.

\item \textit{\textbf{System-level complicated multi-dimensional parallelism and communication modeling.}}
Distributed LLM inference integrates multiple parallelism strategies, including tensor (TP), pipeline (PP), data (DP), expert (EP), sequence (SP), context (CP), and fully sharded data parallelism (FSDP). Each strategy imposes different compute, memory, and communication demands~(\secref{subsubsec:challenge_parallelism}). Existing frameworks rely on simplified linear models~\cite{ko2024dfmodel, zheng2022alpa} or restrict to TP/PP/DP~\cite{astra-simv2, rashidi2025fred},
failing to explore the comprehensive search space.

\item \textit{\textbf{Vast system–hardware co-design space.}}
The interaction between 3D hardware characteristics, thermal constraints, and distributed execution strategies creates a large co-design space spanning per-chip architecture, interconnect topology, parallelism strategy, and execution scheduling~(\secref{subsubsec:challenge_vast_space}).
Optimizing across this space—rather than treating chip- and system-level decisions in isolation—is essential for informed architectural choices.
\end{itemize}

To address these challenges, we propose \oursys{}, a full-stack modeling and DSE framework for distributed 3D-stacked accelerators.
The key distinctions of \oursys{} over prior work
are summarized in Table~\ref{tab:tool_comparison_single_safe}.
At the \textbf{chip level}, \oursys{} performs fine-grained bank-level memory modeling that accounts for data layout, bank activation, traffic distribution, bank conflicts, and transaction-aware bandwidth efficiency.
At the \textbf{system level}, \oursys{} introduces a dual-stage network abstraction combining traffic matrix construction with physical topology mapping and routing, enabling accurate modeling of both on-chip and off-chip interconnects.
This is enhanced by tile-level compute–communication overlap modeling and thermal–power co-modeling that prunes infeasible configurations early.
Built upon these capabilities, \oursys{} supports comprehensive parallelism search across all seven strategies and enables architects to systematically navigate the co-design space to derive \textbf{quantitative design guidance} (\S\ref{ssec:case_study}) for diverse deployment scenarios.
By exposing these coupled decisions in a unified search, \oursys{} provides a shared co-design tool for chip architects and distributed-serving developers. This joint view captures system-scale interactions among network, capacity, and parallelism that roofline analysis alone cannot reveal.
Cross-validated against Cadence Palladium cycle-accurate emulation of our in-house 3D chip and production vLLM serving on 8$\times$B200 GPUs (12.92\% Mean Absolute Percentage Error, or MAPE), \oursys{} achieves consistently low error rates across a wide range of workloads.
Our contributions are summarized as follows:
\begin{itemize}[leftmargin=*]
    \item A full-stack modeling and DSE framework with a 3D-memory model that explicitly maps tiled accesses to DRAM banks, applies a vendor-calibrated transaction-size--bandwidth curve, and enforces Little's-Law buffering constraints. It integrates these chip-level effects with system-level parallelization, execution scheduling, and thermal--power modeling for early-stage design feasibility assessment.
    \item Novel dual-stage network abstraction that reuses each logical traffic-matrix sequence across configurations sharing the same parallelism and collective mapping, while evaluating every synchronization step on the candidate physical topology, together with tile-level compute--communication overlap, achieving up to $100{,}000\times$ speedup over state-of-the-art simulators.
    Cross-validation against Cadence Palladium cycle-accurate emulation and production vLLM serving on 8$\times$B200 (12.92\% MAPE, \S\ref{sssec:h100_ref}) demonstrates modeling fidelity across the full stack.
    \item Comprehensive DSE across ${\sim}2.5\times10^{14}$ configurations identifies designs that achieve up to $9.5\times$ higher modeled throughput than those found from restricted baseline 3D-DRAM design spaces via comprehensive system and hardware co-design.

    \item Beyond modeling and DSE, \oursys{} can be leveraged to provide design guidance (\S\ref{sec:eval}), such as that batch size in distributed settings is a more fundamental architectural driver than the prefill/decode distinction, energy-efficient and throughput-optimized designs require fundamentally different 3D architectures, and incomplete parallelism search misleads distributed 3D design.
\end{itemize}

\input{tables/comparison_v7_single}

%% file: tables/comparison_v7_single.tex
\begin{table}[t]
\centering
\caption{Comparison and novelty over prior work.}
\label{tab:tool_comparison_single_safe}
\footnotesize
\renewcommand{\arraystretch}{1.15}
  \setlength{\tabcolsep}{1.2pt}
  \begin{adjustbox}{max width=\columnwidth}
\begin{tabular}{@{} l | c c c c c c | c @{}}
\toprule
\textbf{Feature}
  & \makecell{\textbf{ASTRA-}\\\textbf{sim v2}\\ \cite{astra-simv2}}
  & \makecell{\textbf{LLM-}\\\textbf{Compass}\\ \cite{llmcompass}}
  & \makecell{\textbf{STCO}\\ \cite{sharda2025system}}
  & \makecell{\textbf{H2-}\\\textbf{LLM}\\ \cite{li2025h2}}
  & \makecell{\textbf{Stra-}\\\textbf{tum}\\ \cite{pan2025stratum}}
  & \makecell{\textbf{Helios}\\ \cite{li2026_3d_Dram_numa_helios}}
  & \makecell{\textbf{Ours} \\ Deep-\\Stack}
\\
\midrule
Fine-Grained 3D Modeling \textsuperscript{1}
   & \ding{55} & \ding{55} & \ding{55} & \ding{55} & \Circle & \Circle & \ding{51}\\
\midrule
Collectives\ Auto Search\textsuperscript{2}
   & \Circle\textsuperscript{3} & \ding{55} & \ding{55} & \ding{55} & \ding{55} & \ding{55} & \ding{51}\\
\midrule
Thermal \& Power
   & \ding{55} & \ding{55} & \ding{51} & \ding{55} & \ding{51} & \ding{51} & \ding{51}\\
\midrule
  \makecell[l]{Distributed Inference: \\ Comprehensive and Flexible \\ Parallelism\textsuperscript{4}}
   & \Circle\textsuperscript{a} & \Circle\textsuperscript{b} & \ding{55} & \ding{55} & \Circle\textsuperscript{c} & \Circle\textsuperscript{d} & \ding{51}\textsuperscript{e}\\
\bottomrule
  \end{tabular}
  \end{adjustbox}\par
\smallskip
\parbox{\columnwidth}{
\scriptsize
\ding{51}: Full support.\ \ \Circle: Partial.\ \ \ding{55}: Not supported.
\\
\textsuperscript{1}Transaction-aware BW, bank activation, Little's Law, bank conflicts.
Stratum models tier-dependent activation latency across monolithic 3D layers; Helios models NUMA-aware PE-bank mapping with per-bank traffic effects. Neither models transaction-size-dependent BW or Little's Law buffering.
\textsuperscript{2}Auto-tuned collective algorithm per topology and message size.\\[2pt]
\textsuperscript{3} ASTRA-sim supports multiple algorithms but user-specified, not auto-searched.\\[2pt]
\textsuperscript{4} Flexible per-module parallelism, e.g., TP in Attention, EP in MoE. \quad
\textbf{Parallelism:}
\textsuperscript{a}TP/PP/DP/FSDP.\
\textsuperscript{b}TP/PP.\
\textsuperscript{c}TP/EP.\
\textsuperscript{d}TP/PP/EP.
\textsuperscript{e}TP/EP/SP/CP/DP/PP/FSDP.\
}
\end{table}

%% file: tex/2_background_v4.tex
\section{Background and Motivation} \label{sec:background}

\subsection{Distributed LLM Inference}
As LLMs scale to hundreds of billions or trillions of parameters~\cite{guo2025deepseek-r1, dubey2024llama3, yang2025qwen3, achiam2023gpt-4},
their footprints exceed single-chip capacity even with stacked DRAM, making \textit{large-scale multi-chip distributed inference} not merely beneficial but essential for serving today's LLMs. A core challenge lies in optimizing parallelism strategies and collective communication patterns, which directly affect hardware utilization.

LLM inference consists of two stages: \textit{prefill}, which processes the prompt and builds the key–value (KV) cache, and \textit{decoding}, which generates tokens sequentially using the KV cache.
Three metrics capture serving performance: \textit{TTFT} (Time to First Token, response latency), \textit{UTPS} (User Tokens Per Second, user-perceived throughput), and \textit{STPS} (System Tokens Per Second, raw serving capacity including speculative or padding tokens).

\subsection{3D-Stacked DRAM Architectures}

\begin{figure}[t]
\includegraphics[width=239pt]{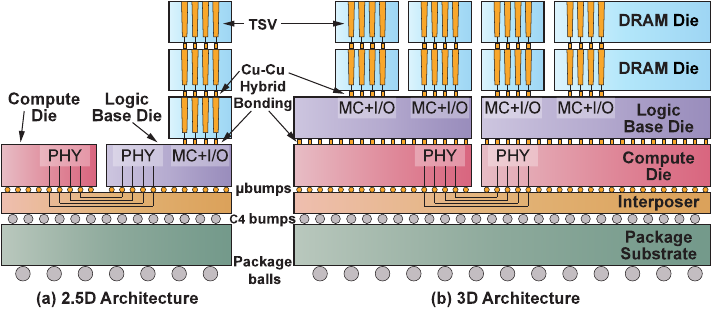}
\caption{2.5D and 3D-stacked architectures (MC: memory controller).}
\vspace{-0.8em}
\label{fig:2d_3d_illustration}
\end{figure}

In conventional 2.5D designs (\figref{fig:2d_3d_illustration}~(a)), compute dies connect to stacked DRAM through an interposer, but bandwidth is constrained by off-stack I/O, and PHY macros occupy tens of $mm^2$~\cite{pan2025stratum}.
As shown in~\figref{fig:2d_3d_illustration}~(b), recent packaging advances enable 3D-stacked accelerators that place DRAM dies directly atop logic dies via TSVs and hybrid bonding, providing high-bandwidth, high-capacity memory close to compute—crucial for LLM decoding. Wafer-scale 3.5D systems~\cite{luomozart} further connect multiple stacks via an interposer for scalable AI.

\subsection{Motivation}

\subsubsection{\textbf{Fine-Grained 3D Hardware Modeling}}\label{subsubsec:challenge_little_law}
\paragraph{\textbf{Little's Law Consideration for 3D-Stacked Accelerators}}
The traditional allocation of area to compute units, on-chip buffers, and network on chip (NoC) resources must be re-evaluated for 3D DRAM-stacked systems.
One of the critical constraints is cache and scratchpad sizing: matching the substantially higher DRAM bandwidth with on-chip storage would require prohibitively large area.
This is closely tied to Little's Law~\cite{little1961proof,nvidia2025gtc_s72683}, where the effective bandwidth achievable is bounded by buffer capacity:
\begin{equation}
\text{Effective Bandwidth} \;\leq\; \text{Buffer Size} \,/\, \text{Latency}
\end{equation}
For instance, the NVIDIA B200 GPU~\cite{nvidia2024blackwell} already requires over $40$\,KiB of shared memory per Thread Block to reach $90\%$ of DDR bandwidth utilization.
However, most prior work~\cite{sharda2025system,llmcompass,pan2025stratum,li2026_3d_Dram_numa_helios} failed to consider this issue, making their modeling questionable for extremely high-bandwidth 3D memory modeling.

\begin{figure}
\includegraphics[width=1.00\linewidth]{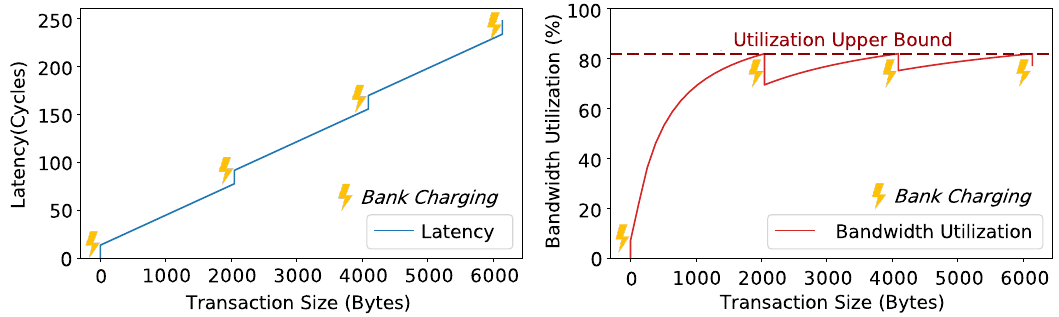}
\caption{Latency and Bandwidth Utilization of 3D-Stacked DRAM (TSV, No Interleaving).}
\vspace{-0.8em}
\label{fig:3d_dram_bw_util}
\end{figure}

\paragraph{\textbf{Unique Bank-Level Access Semantics}}
\label{subsubsec:challenge_dram_transaction_size}
A DRAM die is organized as a set of \emph{banks}: independent memory arrays that can serve requests concurrently, each structured as a 2D array of \emph{rows}. Before
  any data in a row can be accessed, the row must be \emph{activated}, i.e., sensed into the bank's row buffer. After accesses to it complete, the bank must be \emph{precharged} before a different row can be activated. Both
  operations incur fixed latency and energy; requests that alternate between rows within the same bank (\emph{bank conflicts}) serialize on these overheads ~\cite{jacob2010memory,jacob2002dram_tutorial,mutlu2013dram_basics,sinclair2020dram_lecture,rixner2000memory}.
Accurate memory traffic modeling and bandwidth utilization, especially in a distributed 3D system, are dominated by two factors.
First, the most efficient pattern is to \emph{stream a full bank row (e.g., $\sim$2\,KiB)}. As shown in Fig.~\ref{fig:3d_dram_bw_util}, partial accesses fail to amortize activation and precharge overheads, leading to lower effective bandwidth.
Although techniques such as customized DRAM arrays or sub-row activation can mitigate this~\cite{o2017fine}, they introduce additional design complexity and area/power overhead.
Second, stacked DRAM exposes a large number of independently connected banks (tens to hundreds). Achieving peak bandwidth requires concurrent streaming across many banks, so memory traffic smaller than a few hundred kilobytes generally cannot saturate TSV bandwidth.

Consequently, when per-bank traffic is small, e.g., under fine-grained tiling or aggressive parallel sharding, achieved bandwidth drops sharply. However, most previous modeling works~\cite{llmcompass,sharda2025system,li2025h2} assume idealized bank interleaving or uniform streaming and thus \textbf{overestimate} achievable bandwidth in practical distributed LLM serving.

\paragraph{\textbf{Modeling of 3D-Stacked Connectivity}} To maximize DRAM bandwidth, 3D-stacked accelerators can connect each bank directly to the logic die via TSVs, bypassing traditional multi-bank interleaving used to hide activation and turnaround latency~\cite{rixner2000memory,jacob2010memory,hadidi2017demystifying}.
However, direct bank access is only one point in the design space: practical systems may also instead employ multi-bank interleaving or connect only a subset of the stacked layers.
To facilitate the exploration of this design choice,
\oursys{} supports both direct-access and interleaved designs, where we parameterize both modes and investigate their impact in \S~\ref{ssec:connected_vs_stacked}.

\subsubsection{\textbf{Compound Parallelism Strategies}}\label{subsubsec:challenge_parallelism}
\begin{figure}[t]
    \centering
    \includegraphics[width=0.8\linewidth]{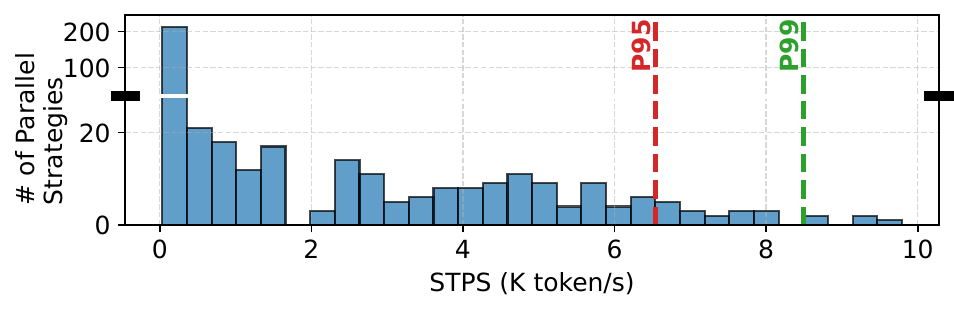}
    \caption{
    The number of distinct parallelisms achieved at a certain STPS, showing a wide performance distribution.
    }
    \label{fig:distribution}
    \vspace{-1em}
\end{figure}

Distributed LLM serving combines multiple parallelism strategies, including TP, PP, DP, EP, SP, CP, and FSDP~\cite{svedas2025survey, kwon2023vllm, zheng2024sglang}.
Each  strategy leads to different per-device compute, memory, and communication characteristics.
As a result, the optimal parallelism combination usually varies from model to model.
As shown in Fig.~\ref{fig:distribution}, performance varies significantly across different parallelism configurations for a given model, and high-performance strategies are rare.
However, existing frameworks typically use simplified cost models~\cite{ko2024dfmodel, zheng2022alpa, astra-simv2} or restrict parallelism to TP/PP/DP~\cite{astra-simv2, rashidi2025fred}, without accurate modeling of compute–communication overlap at tile granularity. For instance, limited parallelism strategies force mixture of experts (MoE) models to shard experts via large TP, reducing throughput by up to $5\times$ (\S\ref{ssec:ablation}).
An incomplete search space can mislead the DSE into incorrect chip designs irrecoverable after fabrication.

\subsubsection{\textbf{Vast Co-Design Space and Thermal Constraints}}\label{subsubsec:challenge_vast_space}

\paragraph{\textbf{System–Hardware Co-Design}}
A comprehensive system and hardware co-design requires exploring broad design choices:
\begin{itemize}[leftmargin=*]
\item \textbf{Per-chip architectural configuration}: compute units, on-chip buffering, and memory hierarchy.
\item \textbf{On-chip and inter-chip network topology}: bandwidth provisioning and latency/area trade-offs.
\item \textbf{Distributed parallelism planning}: selection and mapping of TP/PP/DP/EP/SP/CP to physical devices.
\item \textbf{Single-chip scheduling}: tile size, kernel fusion, memory placement, and scheduling pipelining.
\item \textbf{Collective communication strategies}: choice of algorithms (e.g., all-reduce versus reduce-scatter+all-gather) and fine-grained compute–communication overlap.
\end{itemize}
More critically, these dimensions are tightly coupled: parallelism strategies determine per-chip memory traffic, which in turn dictates buffer sizing (via Little's Law) and optimal DRAM stacking depth. Yet prior approaches typically optimize them independently or explore only a subset. As shown in \S\ref{ssec:ablation}, our joint DRAM-layer and NoC co-optimization contributes 33--39\% throughput improvement over a fixed configuration.
\paragraph{\textbf{Thermal Consideration}}
The thermal issue further compounds the co-design space: additional DRAM layers increase thermal resistance, and as we demonstrate in \S\ref{ssec:case_study}, many high-bandwidth configurations exceed the 85$^\circ$C thermal limit under sustained decode workloads. Without thermal-aware modeling, architects risk committing to configurations that appear optimal in performance models but are thermally infeasible in practice.

%% file: tex/3_design_v3.tex
\section{\oursys{} Framework} \label{sec:design}

\subsection{\oursys{} Overview}
\label{ssec:workflow}

\begin{figure*}[t]
\centering
\includegraphics[width=1.00\linewidth]{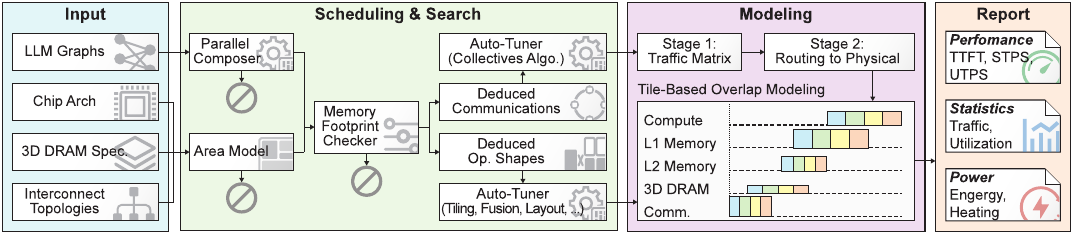}
\vspace{-1em}
\caption{Overview of \oursys{} DSE Framework. }
\vspace{-0.5em}
\label{fig:dse_framework}
\end{figure*}

\begin{figure*}
\includegraphics[width=515.52pt]{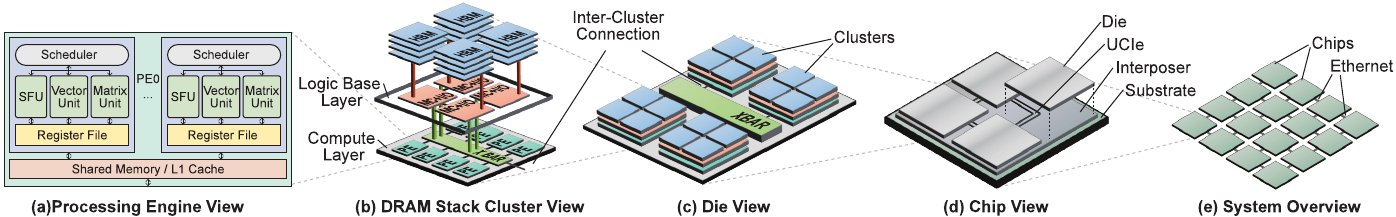}
\vspace{-0.5em}
\caption{Example of Cross Sectional and Top View of a 3D-Stacked DRAM Architecture.}
\label{fig:hierarchical_hardware}
\vspace{-1em}
\end{figure*}

\figref{fig:dse_framework} illustrates an overview of \oursys{}.
It takes a model computation graph supporting diverse LLM building blocks,
along with batch size, sequence lengths, and decoding modes.
On the hardware side, it consumes candidate configurations of compute-die parameters (buffer sizes, compute units) and NoC settings (hierarchical topologies with bandwidth and latency). All hardware configurations are first validated through an area cost model~(\secref{subsec:area_model}) given a per-die area budget.

For feasible hardware, a parallel-strategy composer enumerates configurations across TP/EP/SP/CP/DP/FSDP/PP.
An empirical filter removes invalid or non-beneficial options (e.g., SP for single-step decoding, FSDP when DP$=1$).
A memory-footprint checker then prunes configurations whose model size, KV-cache, and peak activations exceed memory capacity.

For each remaining configuration, \oursys{} determines per-chip operator shapes and collective communication primitives. Two auto-tuners operate at this stage:
(i) an operator-level tuner searching tiling, fusion, and layout strategies for wave-level execution, and
(ii) a collective tuner selecting the most efficient algorithm for the given topology and data size.

Finally, a tile-based compute–communication performance model (\secref{sssec:overlap_modeling}) simulates the full execution timeline while maximizing overlap.
It produces a report with TTFT, UTPS, STPS, and utilization statistics across compute, cache, stacked DRAM, and the hierarchical NoC, guiding further architectural refinement.

\subsection{Hardware Model}
\oursys{} uses a hierarchical hardware model instantiated from our in-house 3D designs. In the configuration studied here, each cluster's PEs directly access the DRAM banks stacked above the cluster, while the cluster size and stacking depth remain configurable.
As shown in Fig.~\ref{fig:hierarchical_hardware}, it progresses from a fine-grained processing engine to the full multi-chip system, which consists of:

\textbf{(a) Processing Engine (PE).}
The PE is the fundamental compute unit, analogous to a Streaming Multiprocessor of GPUs.
Each PE (\figref{fig:hierarchical_hardware}(a)) contains a configurable mix of Special Function Units (SFUs), Vector units, and Matrix units, along with an L0 register file and L1 shared memory with configurable capacity and bandwidth.
Users can also specify the minimum matrix tile shape.

\textbf{(b) 3D-Stacked DRAM Cluster.}
As shown in~\figref{fig:hierarchical_hardware}(b), a cluster consists of multiple PEs, an L2 cache, local interconnect, and vertically stacked DRAM layers bonded over a logic base die.
Configurable parameters include PE count, L2 size/bandwidth, DRAM layer count (e.g., 4-layer stack), and bank size (typically 2\,KiB).
\oursys{} distinguishes stacked layers (contributing capacity) from connected layers (additionally contributing bandwidth).
An example of architectural options is shown in Table~\ref{tab:arch_config}.

\textbf{(c) Die Level.}
A die (\figref{fig:hierarchical_hardware}(c)) contains multiple clusters connected via a configurable Level-1 (L1) network with selectable topology, link bandwidth, and per-hop latency.

\textbf{(d) Chip Level.}
A chip (\figref{fig:hierarchical_hardware}(d)) integrates multiple dies via universal chiplet interconnect express (UCIe) links, modeled with a Level-2 (L2) network with configurable topology and link parameters.

\textbf{(e) System Level.}
Multiple chips are linked via Ethernet (\figref{fig:hierarchical_hardware}(e)), forming the Level-3 (L3) network with configurable topology and bandwidth.

For \textbf{(c)--(e)}, \oursys{} supports six fundamental topologies: \emph{switch}, \emph{1D chain/ring}, \emph{2D mesh/torus}, and \emph{fully connected}, with hierarchical compositions across L1/L2/L3.
Users can specify routing rules and port mappings for hybrid topologies.
Example network configurations are in Table~\ref{tab:noc_settings}.

\subsection{System Scheduling Configuration}

System performance is highly sensitive to scheduling, especially in distributed settings. Therefore, \oursys{} incorporates comprehensive distributed parallelisms and collective communication choices into the design space.

\subsubsection{Parallel Strategy}

\oursys{} supports a wide range of parallelism combinations, including TP~\cite{shoeybi2019megatron},
EP~\cite{lepikhin2020gshard,fedus2022switch}, SP~\cite{li2023sequence}, CP~\cite{Megatron-LM}, DP~\cite{krizhevsky2014one}, FSDP~\cite{zhao2023pytorch},
and PP~\cite{huang2019gpipe, narayanan2019pipedream}.
Following the common practice from Megatron~\cite{shoeybi2019megatron} and DeepSpeed~\cite{rasley2020deepspeed}:
\textbf{TP} splits attention heads and FFN/MoE intermediate dimensions,
\textbf{EP} partitions routed experts (defaulting to DP for non-MoE layers),
\textbf{SP} partitions the sequence dimension, mainly in prefill,
\textbf{CP} splits KV-cache across devices for both prefill and decode,
\textbf{PP} partitions layers with microbatching for pipeline efficiency,
\textbf{DP} splits the batch, and \textbf{FSDP} further shards weights on top of DP.

Given a target parallelism degree~$N$, the composer enumerates all valid, non-duplicate factorizations:
\begin{equation}
    TP \times EP \times SP \times CP \times DP \times PP = N.
\end{equation}
Candidates are enumerated by prime-factorizing $N$ and applying stars-and-bars allocation across six dimensions. Since FSDP is optional with DP, each candidate can be set with \texttt{FSDP=True} or \texttt{False}.

\subsubsection{Single Chip Scheduling}
Single-chip scheduling has been extensively explored in prior compilers and frameworks~\cite{chen2018tvm,zhu2022roller,shi2023welder,wang2024ladder,cheng2025pipethreader,osama2023stream,cutlass,tilelang}. \oursys{} incorporates scheduling spaces, including tiling, fusion, and software pipelining to hide memory latency.
For instance, in matrix multiplication, it  searches over tile sizes across memory hierarchies, the number of software-pipeline stages to overlap computation with memory access, and fusing lightweight epilogue operations such as bias addition.

Beyond classical optimizations, \oursys{} adds constraints from 3D-stacked DRAM: directly connected banks without interleaving can cause wasted accesses or poor row locality. We incorporate penalty terms for strategies that lower effective bank utilization or cause bank conflicts due to layout misalignment.

\subsubsection{Collectives Implementation}
Once a parallel strategy is fixed, the required communication collectives for tensor reconstruction are determined. However, each collective operation admits multiple algorithmic implementations, whose performance varies widely depending on the underlying network topology, link bandwidth, and hop latency.
For instance, all-reduce can be implemented using ring, Rabenseifner, double-tree, and double-recursive halving-and-doubling algorithms.
Our \oursys{} evaluates mainstream candidate algorithms for each collective, and the modeling stage selects the highest-performing option based on topology-aware communication costs and potential computation overlap.

\section{Modeling and Design Space Exploration}
\subsection{Modeling Methodology}
\subsubsection{Area Modeling}\label{subsec:area_model}
We synthesized our baseline design ("Standard" in \tabref{tab:arch_config}, "TMS1" in \tabref{tab:noc_settings}) using the ASAP7 7\,nm predictive PDK~\cite{clark2016asap7} and extracted post-synthesis area figures as reference anchors. During exploration, compute-unit areas are linearly scaled by throughput~\cite{llmcompass, li2025h2}. On-chip SRAM areas are decomposed into capacity-proportional and bandwidth-proportional parts, calibrated from Memory Compiler-generated macros~\cite{pan2025stratum}. Memory controller area scales with peak DRAM bandwidth based on~\cite{xie2025memory_controller}, with additional DRAM peripheral overhead derived from published logic-die breakdowns~\cite{li2025h2, pan2025stratum}. On-chip interconnect area is scaled from the baseline with sub-linear bandwidth elasticity. Total die area includes a 15\% overhead for control and routing, observed from our baseline synthesis. Designs exceeding the area budget are pruned before performance evaluation.

\subsubsection{Power and Thermal Modeling}\label{subsec:power_model}
In 3D stacking, overheating increases leakage and can render the device non-functional. Typically, 95\,\textdegree C is the hard limit~\cite{sharda2025system} and 85\,\textdegree C is a safer target~\cite{yue20253d-path, yue2024exploiting, li2026_3d_Dram_numa_helios}.
\paragraph{\textbf{Power Model.}}
\oursys{} collects per-operator memory traffic (register, shared memory, L2, DRAM), NoC traffic, and compute operations (vector, SFU, matrix), then multiplies by technology-calibrated per-bit energy coefficients for memory~\cite{horowitz20141} and NoC~\cite{pasricha2020survey, mota2022ucie, li2018112, marcon2005exploring}, as well as per-operation energy for compute~\cite{o2017fine, wu2019seco, mach2020fpnew}. Static power is 10\% of Thermal Design Power (TDP)~\cite{o2017fine}.
\paragraph{\textbf{Thermal Model.}}
Junction temperature follows $T = T_{\text{amb}} + R(m) \cdot P$. We anchor at a 4-layer, 100\,W baseline with 6000\,W/m$^2$K convection, and sweep $m \in [1,12]$ in ANSYS Fluent~\cite{lee2025thermal_resistance} to fit a linear model: $R(m) = 0.56 + 0.01 \cdot m$ (\textdegree C/W). We cross-check this model against HotSpot~\cite{huang2006hotspot,han2021_hotspot7} in~\secref{subsec:model_acc}.
\paragraph{\textbf{Dynamic Voltage and Frequency Scaling (DVFS)}}
More DRAM layers raise $R(m)$, reducing sustainable power. Since $P_{\text{dyn}} \propto V^2 f$ and $V \propto f$~\cite{horowitz20141}, we have $P_{\text{dyn}} \propto f^3$. The thermal-limited frequency scales as $f_{\text{th}} = ({P_{\text{dyn}}(m)}/{P_{\text{dyn}}(4)})^{1/3}$, where $P_{\text{dyn}}(m) = \text{TDP} \times R(4)/R(m) - P_{\text{static}}$. A second \emph{power-wall} check further reduces frequency if actual per-device power exceeds TDP.

\subsubsection{Single-Die Performance Modeling}\label{sssec:model_3d_dram}
Our single-die model follows a tile-level abstraction inspired by prior frameworks~\cite{parashar2019timeloop,kwon2020maestro,zheng2023tileflow,huang2024mind, mo2025lut, lee2025forecasting}, which have shown tile-level simulation is sufficiently accurate for non-3D architectures.
To incorporate the unique behaviors of 3D-stacked DRAM, we identify four key features:
\begin{enumerate}[label=(\roman*), leftmargin=*]
    \item \textbf{Transaction-size–dependent bandwidth.}
    Without multi-bank interleaving, peak bandwidth requires reading a whole bank row per access. Achievable bandwidth decreases with smaller transactions.
    \item \textbf{Constrained buffering imposed by Little's Law.} Saturating TSV bandwidth requires sufficient buffering ($\text{buffer} \ge \text{bandwidth} \times \text{latency}$). Insufficient buffering directly reduces throughput.
    \item \textbf{Bank parallelism limited by operator size.}
    With directly attached banks, small operators may not issue enough requests to activate all banks, reducing effective bandwidth.
    \item \textbf{Bank conflicts from layout and access patterns.}
    Without interleaving, certain tensor layouts may map many tiles to the same bank, introducing conflict-induced serialization.
\end{enumerate}

To model these efficiently, \oursys{} applies the following rules.
For \textbf{(i)}, we obtain achievable bandwidth from the bandwidth\hspace{0pt}–transaction-size curve in~\figref{fig:3d_dram_bw_util}, derived from the 3D-DRAM vendor's C-model. For each tile access, our method computes the effective transaction size and queries this curve.
For \textbf{(ii)}, bandwidth is further bounded by Little's Law:
\begin{equation}
\mathrm{BW}_{\mathrm{eff}} \;=\; \min\!\left(\mathrm{BW},\;\text{buffer} \,/\, \text{latency}\right)
\end{equation}
For \textbf{(iii)} and \textbf{(iv)}, \oursys{} explicitly maps each tensor tile to a DRAM bank according to the chosen layout and swizzling policy. For every memory wave, we build a per-bank access histogram.
The most heavily loaded bank determines the wave's service time. This unified mechanism simultaneously models \textit{(a)} limited bank-level parallelism for small operators and \textit{(b)} bank conflicts caused by
less efficient data layouts.
Based on these, \oursys{} preserves the efficiency of tile-level
modeling while capturing the core performance-shaping behaviors unique to
3D-stacked DRAM architectures.

\subsubsection{On-Chip/Off-Chip Network Modeling}
\label{sssec:noc_modeling}
The primary goal of our network modeling is to efficiently estimate latency and link utilization across diverse topologies.
Existing tools such as ASTRA-sim-v2~\cite{astra-simv2} face significant trade-offs: analytical backends can incur up to 58\% error, while NS-3~\cite{riley2010ns-3} backends require over an hour per GiB-scale collective, making it infeasible for billion-point DSE.
To address this challenge, we propose a novel dual-stage network abstraction, which is described as follows:

\textbf{Stage-1: Traffic Matrix Construction.}
We represent each collective communication step using a Traffic Matrix (TM), a logical $\text{num\_nodes} \times \text{num\_nodes}$ matrix where entry $(s,d)$ denotes the data volume sent from source $s$ to destination $d$.
All traffic within a TM starts simultaneously, while complex collectives decompose into TM sequences. Crucially, the TM captures logical patterns independent of physical topology. During construction, \oursys{} aggregates traffic from all parallelism dimensions.
To reflect realistic optimization, \oursys{} prioritizes mapping logically intensive communication pairs (e.g., TP) to physically adjacent nodes. This ensures that high-volume logical edges benefit from higher bandwidth and lower latency in the physical layer.

\begin{figure}[t]
    \centering

    \subfloat[Traffic Matrix.\label{fig:tm_logical}]{%
        \includegraphics[width=0.48\linewidth]{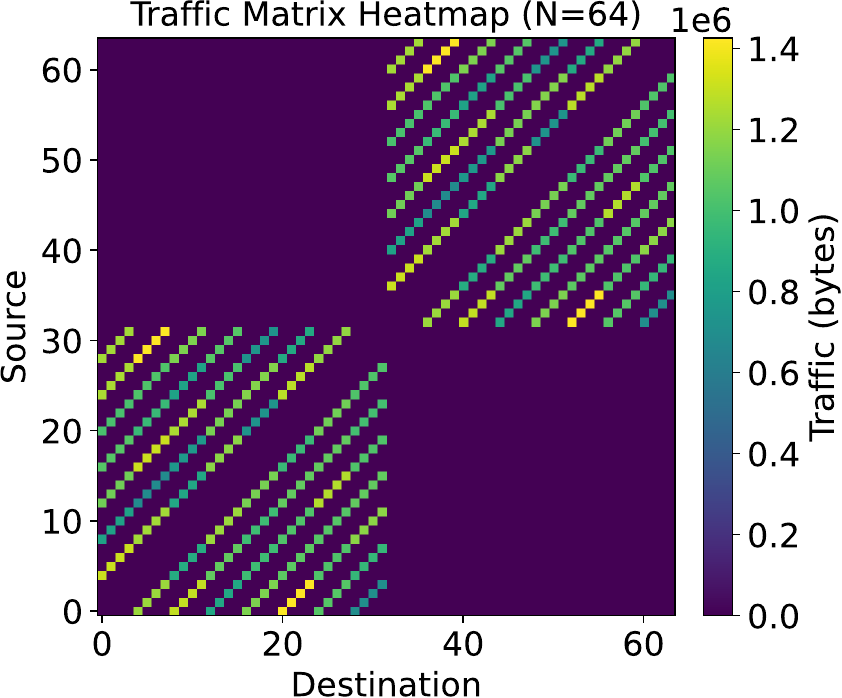}}
    \hfill
    \subfloat[Topology bandwidth.]{%
        \includegraphics[width=0.48\linewidth]{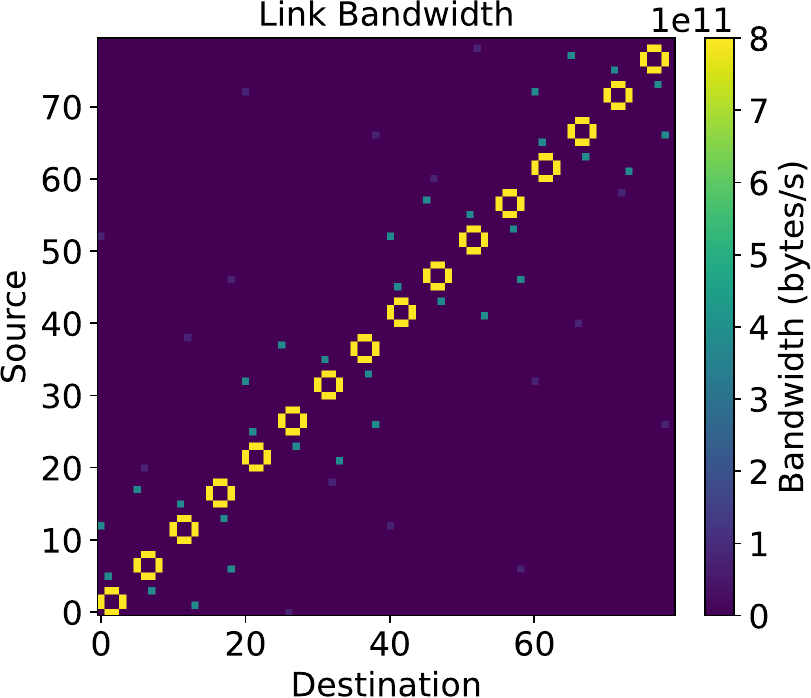}}

    \subfloat[Traffic after routing.]{%
        \includegraphics[width=0.48\linewidth]{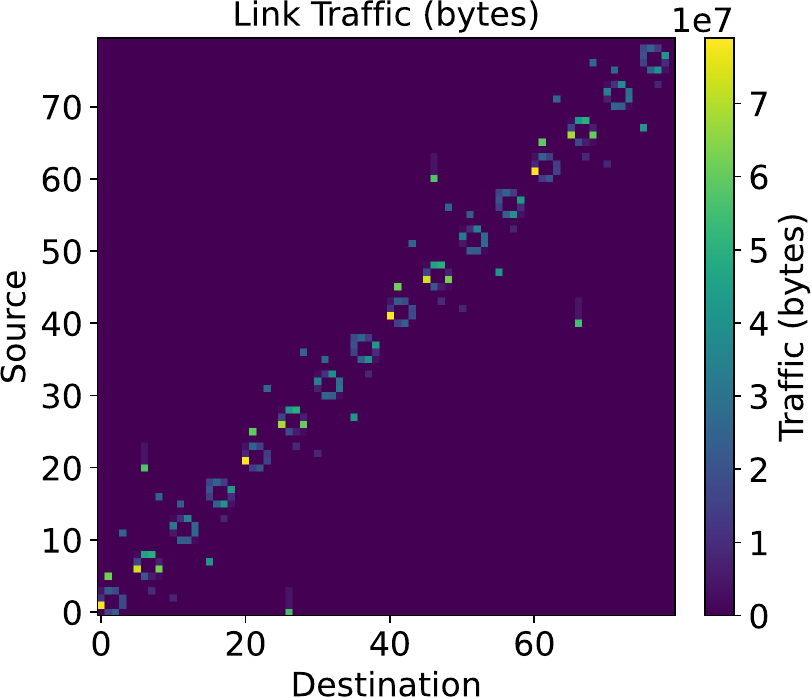}}
    \hfill
    \subfloat[Final utilization.]{%
        \includegraphics[width=0.48\linewidth]{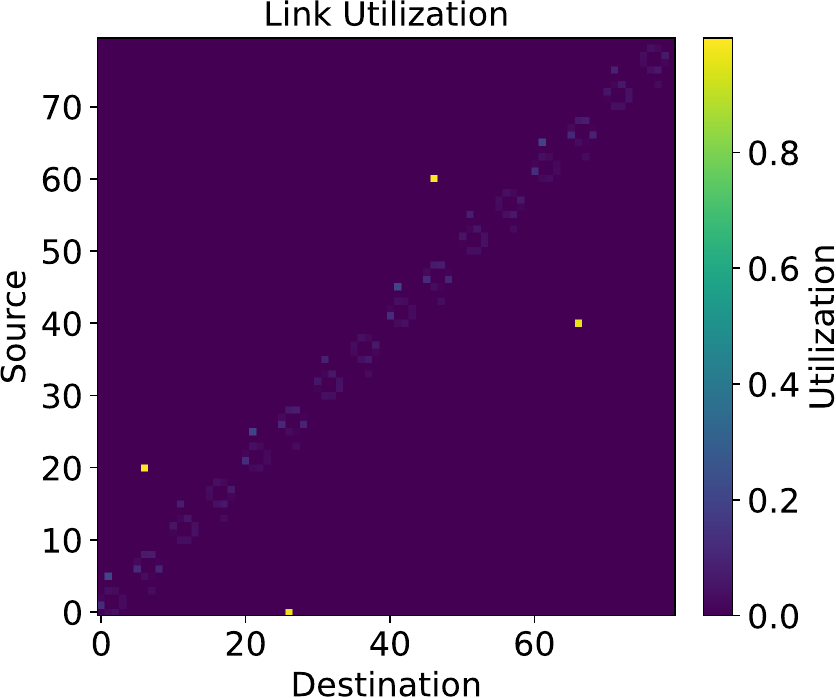}}

    \caption{
    Example of mapping 64-node logical EP all-to-all traffic onto a three-layer torus–mesh–mesh topology. (a) Logical TM, (b) topology bandwidth, (c) routed physical traffic, and (d) resulting link utilization.
    }
    \vspace{-0.5em}
    \label{fig:traffic_matrix_mapping}
\end{figure}

\textbf{Stage-2: Physical Mapping and Routing.}
Given a TM, \oursys{} maps logical $(s,d)$ flows to physical routes defined by the target topology.
After routing, it accumulates the traffic volume $V_\ell$ on each physical link $\ell$ with bandwidth $BW_\ell$.
The total network time $T_{net}$ is modeled as the sum of the structural hop latency and the bottleneck serialization delay:
\begin{equation}\label{eq:network_time}
T_{net}
= \underbrace{\max_{p \in \mathcal{P}} \left(\sum_{\ell \in p}\delta_{\ell}\right)}_{L_{net}}
\;+\; \underbrace{\max_{\ell \in \mathcal{L}} \left(\frac{V_\ell}{BW_\ell}\right)}_{T_{cong}},
\end{equation}
where $\mathcal{P}$ is the set of active paths, $\mathcal{L}$ is the set of physical links, and $\delta_\ell$ is the zero-load latency of link $\ell$.
The first term $L_{net}$ is the maximum accumulated zero-load latency over all active paths, used as a key parameter in overlap modeling (\S~\ref{sssec:overlap_modeling}).
The second term $T_{cong}$ captures volume-based contention through the bottleneck link: flows sharing a physical link increase $V_\ell$ and hence its serialization time.
Rather than applying the standard \mbox{$\alpha$--$\beta$} cost model~\cite{thakur2005optimization} directly to an aggregate message, our two-stage representation first constructs a sequence of logical traffic matrices for each collective and then maps every synchronization step onto the physical topology. For configurations sharing the same parallelism and collective mapping, the logical sequence is reused while physical mapping and routing are reevaluated, improving DSE efficiency without discarding step-level communication structure.
Fig.~\ref{fig:traffic_matrix_mapping} visualizes this process of mapping the traffic matrix to the final utilization based on physical-topology bandwidth.
Although dynamic queueing and finite-buffer backpressure are not modeled, this two-stage abstraction still captures volume-based link contention without event-level simulation. It reduces evaluation time to under 0.1 seconds for a 256-node topology while maintaining less than 5\% error compared to an NS-3 backend (Fig.~\ref{fig:noc_modeling_accuracy}), making billion-scale DSE feasible.

\subsubsection{Compute-Communication Overlap Modeling}\label{sssec:overlap_modeling}
\begin{figure}
\includegraphics[width=1.00\linewidth]{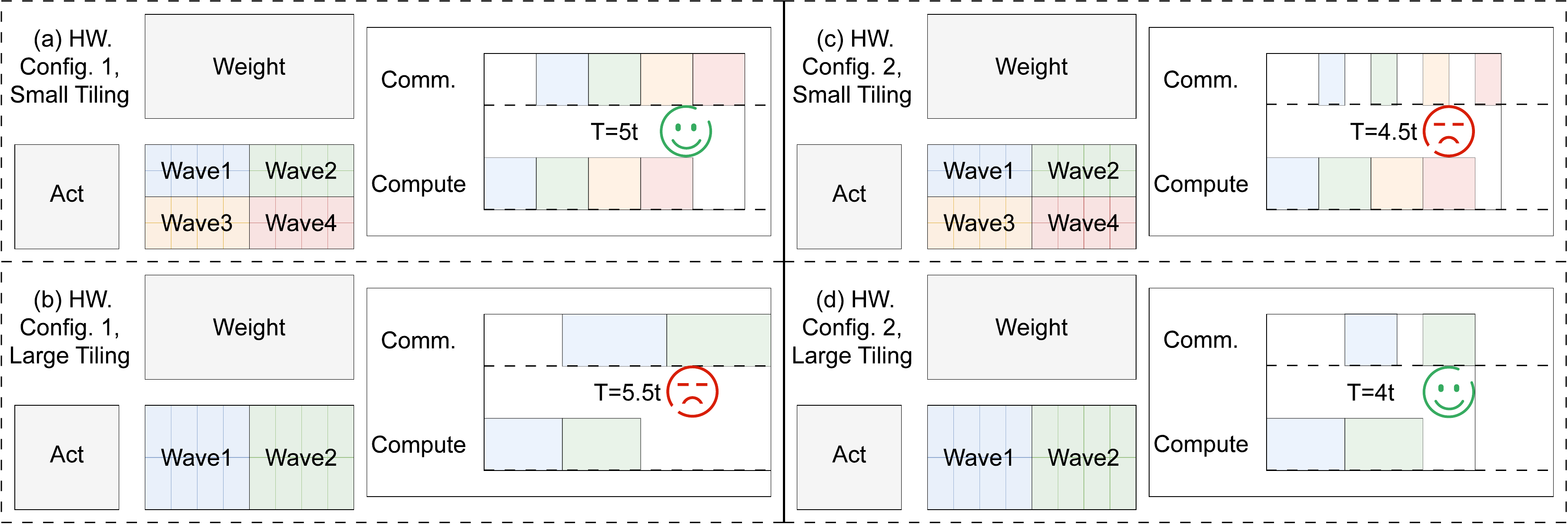}
\caption{
\textbf{Tile-level} compute–communication overlap modeling for GEMM.
Panels compare Low (a,b) vs. High (c,d) NoC bandwidths using Small (a,c) vs. Large (b,d) tiling strategies.
The visualization demonstrates how optimal tiling shifts from overlap-oriented (small tiles) to compute-oriented (large tiles) as network bandwidth increases.
}
\vspace{-0.5em}
\label{fig:compute_communication_overlap}
\end{figure}

To facilitate accurate modeling of distributed LLM inference, \oursys{} models compute--communication overlap at the granularity of tile-based wavefronts. As illustrated in Fig.~\ref{fig:compute_communication_overlap}, once the computation for a wave of tiles is completed, \oursys{} initiates the corresponding data transfer. This allows the communication of the $i$-th wave to proceed in parallel with the computation of the $(i+1)$-th wave.

Let $W = \lceil \text{num\_tiles} / \text{num\_SMs} \rceil$ denote the number of waves. We define the per-wave compute time as $\tau_{comp}$ and the per-wave transmission time as $\tau_{comm}$.
For $W > 1$, the end-to-end latency $T_{e2e}$ is modeled as a three-stage pipeline, incorporating the network hop latency $L_{net}$ derived in \S~\ref{sssec:noc_modeling}:
\begin{equation}
\label{eq:overlap_model}
\begin{split}
    T_{e2e} &= \underbrace{\tau_{comp}}_{\text{Prologue}} + \underbrace{(\tau_{comm} + L_{net})}_{\text{Epilogue}} \\
            &\quad + \max\bigl( (W\!-\!1)\tau_{comp}, (W\!-\!1)\tau_{comm} + L_{net} \bigr)
\end{split}
\end{equation}

The term in the second line represents the steady-state pipelined phase. Note that $L_{net}$ is explicitly included in the communication terms to account for the signal propagation delay that cannot be hidden by bandwidth-bound pipelining.
If $W \le 1$, the model reduces to a serialized execution: $T_{comp} + W \cdot \tau_{comm} + L_{net}$.
This modeling formulation highlights a critical architectural trade-off between compute efficiency and overlap efficiency.
Larger tile sizes generally improve data reuse (higher $\tau_{comp}$ efficiency) but result in fewer total waves ($W$). A smaller $W$ reduces the depth of the pipeline, making the non-overlappable prologue and epilogue a larger fraction of the total execution time.
Conversely, smaller tiles increase $W$, maximizing overlap opportunities, but may degrade arithmetic intensity.
As shown in Fig.~\ref{fig:compute_communication_overlap}, with low NoC bandwidth (panels a, b), small tiling (a) is preferable to maximize waves and hide communication. However, with high NoC bandwidth (panels c, d), the communication cost keeps low, making large tiling more favorable since it prioritizes compute efficiency over overlap. \oursys{} automatically navigates this complex design space to find the optimal configuration.

\subsection{DSE Framework}
\subsubsection{Workloads}
\oursys{} supports a broad range of LLM architectures by operating directly on their computation graphs. For attention mechanisms, it includes standard Multi-Head Attention (MHA)~\cite{vaswani2017attention}, Grouped-Query Attention (GQA)~\cite{ainslie2023gqa}, Multi-Query Attention (MQA)~\cite{chowdhery2023palm}, and DeepSeek's Multi-Head Latent Attention (MLA) with automatic absorbed/non-absorbed form selection. For feed-forward networks (FFNs), both dense SwiGLU layers (e.g., LLaMA family) and Mixture of Experts (MoE) blocks (e.g., DeepSeek, Qwen3) are supported. To ensure accurate MoE communication and load-balance modeling, we extract realistic routing patterns from actual inference traces.

\subsubsection{Design Space Size}

\label{sssec:searching_space}

As summarized in the workflow (\S~\ref{ssec:workflow}), the full design space spans both hardware and software dimensions. At the hardware level, \oursys{} explores single-chip architectural configurations, 3D DRAM stacking depth ($m{=}1$--$16$ stacked layers with $n{\leq}m$ connected layers), on-chip SRAM capacity/bandwidth settings, on-chip/off-chip network topologies, NoC hop latency, and per-layer NoC bandwidth, all subject to area and thermal constraints. Given a feasible hardware configuration, the system level introduces three additional degrees of freedom: distributed parallel strategies, single-chip operator scheduling, and collective-communication algorithm selection. Together, these form a vast joint hardware--system search space.
For any given configuration, \oursys{} generates an end-to-end performance report including TTFT, UTPS, STPS, as well as detailed utilization statistics for memory, compute, and each NoC layer, along with power and thermal estimates~(\S~\ref{subsec:power_model}).
\input{tables/arch_settings}
\input{tables/noc_settings}
However, this raw design space is extremely large.  Consider a scenario with batch sizes $\{1,16,128,1024\}$, six input/output sequence lengths, and 4 model architectures. On the hardware side, we explore 30 base architecture/topology combinations, 136 stacking configurations ($m{=}1$--$16$, $n{\leq}m$), 3 SRAM capacity/bandwidth settings, 5 NoC hop-latency multipliers, and per-layer NoC bandwidth scaling across the three-level hierarchy ($4^3{=}64$ combinations). With 256 devices, the software space comprises 2{,}574 parallel strategies, 4 collective algorithms, and ${\sim}$64 single-chip schedules ($6.6 \times 10^5$ per setting). The total theoretical design space reaches ${\sim}2.5\times 10^{14}$ configurations.

\subsubsection{Design Space Pruning and Hierarchical Search.}
\label{sssec:pruning}
To enable efficient search on the vast design space (up to ${\sim}2.5\times 10^{14}$), we apply multi-stage pruning to make this space tractable.
\emph{(1)~Pareto-dominated hardware pruning:} Configurations where stacked layers $m{>}10$ are eliminated, as Little's law and L1 bandwidth constraints cause effective DRAM bandwidth to \emph{decrease} beyond this point despite increasing theoretical bandwidth (\S\ref{ssec:case_study}), making them strictly dominated on both bandwidth and compute FLOPS.
\emph{(2)~Parallel-strategy pruning:} Invalid or dominated strategies are removed (e.g., $SP{>}1$ in single-step decoding, $FSDP{=}\text{True}$ with $DP{=}1$, infeasible $DP/PP$ given batch size), eliminating ${\sim}$80\% of candidates.
\emph{(3)~Memory-footprint check:} Model weights, KV cache, and peak activations must fit in DRAM (with 10\% reserved for CUDA graphs and fragmentation), pruning ${\sim}$50\% of remaining configurations.
\emph{(4)~Hierarchical NoC search:} Rather than exhaustively enumerating the $5{\times}64{=}320$-point NoC latency/bandwidth grid, we adopt a hierarchical approach. \oursys{} first searches over base architecture and stacking configurations, then advances only the top 5\% into the NoC latency sweep, and finally promotes the top 5\% of that tier into per-layer bandwidth tuning.
Together, these strategies reduce the run to approximately \mbox{$1\times10^{8}$} evaluated configurations, completed in about two days on a 512-core CPU.

%% file: tables/arch_settings.tex
\begin{table}[t]
\caption{Architecture Specifications per $800\,\text{mm}^2$ }
\centering
\scalebox{0.85}{
\begin{threeparttable}
\setlength{\tabcolsep}{2pt}
\begin{tabular}{rcccccc}
\hline
                      & \textbf{\begin{tabular}[c]{@{}c@{}}FP16\\ TFLOPS\end{tabular}} & \textbf{\begin{tabular}[c]{@{}c@{}}FP32\\ TFLOPS\end{tabular}} & \textbf{\begin{tabular}[c]{@{}c@{}}SFU\\ TFLOPS\end{tabular}} & \textbf{\begin{tabular}[c]{@{}c@{}}L2BW\\ (TB/s)\end{tabular}} & \textbf{\begin{tabular}[c]{@{}c@{}}L1BW\\ (TB/s)\end{tabular}} & \textbf{\begin{tabular}[c]{@{}c@{}}WG\\ MMA\end{tabular}} \\ \hline
\textbf{H100*}         & 504.6                                                          & 31.5                                                           & 2.0                                                           & 9.8                                                            & 15.8                                                              & \ding{52}                                                 \\
\textbf{H200*}         & 504.6                                                          & 31.5                                                           & 2.0                                                           & 11.8                                                           & 15.8                                                              & \ding{52}                                                 \\
\textbf{Standard}     & 367.0                                                          & 11.5                                                           & 1.4                                                           & 13.1                                                           & 22.9                                                              & \ding{56}                                                 \\
\textbf{WGMMA}        & 367.0                                                          & 11.5                                                           & 1.4                                                           & 13.1                                                           & 22.9                                                              & \ding{52}                                                 \\
\textbf{Weak NoC}     & 412.9                                                          & 12.9                                                           & 1.6                                                           & 13.1                                                           & 25.8                                                              & \ding{56}                                                 \\
\textbf{Strong NoC}   & 321.1                                                          & 10.0                                                           & 1.3                                                           & 13.1                                                           & 20.0                                                              & \ding{56}                                                 \\
\textbf{Strong L2}    & 275.3                                                          & 8.6                                                            & 1.1                                                           & 32.0                                                           & 17.2                                                              & \ding{56}                                                 \\
\textbf{Strong L1}    & 321.1                                                          & 10.0                                                           & 1.3                                                           & 13.1                                                           & 40.1                                                              & \ding{56}                                                 \\
\textbf{Less SM}      & 321.1                                                          & 10.0                                                           & 1.3                                                           & 13.1                                                           & 20.1                                                              & \ding{56}                                                 \\
\textbf{Large Matrix}     & 458.8                                                          & 14.3                                                           & 1.8                                                           & 13.1                                                           & 28.7                                                              & \ding{56}                                                 \\
\textbf{Large Vector} & 321.1                                                          & 20.1                                                           & 2.5                                                           & 13.1                                                           & 20.1                                                              & \ding{56}                                                 \\ \hline
\end{tabular}
    \begin{tablenotes}
        \footnotesize
        \item H100* and H200* denote scaling them from 4\,nm to 7\,nm to decrease SM counts while keeping other aspects unchanged.
        \item DRAM configuration: H100*: 3.3\,TB/s, 80\,GiB HBM3;  H200*: 4.8\,TB/s, 141\,GiB HBM3e;  Others (4-layer 3D-stacked): 13.1\,TB/s, 64\,GiB.
        \item WGMMA denotes support for four matrix units sharing input operands.
        \item All configurations satisfy the 800\,mm\textsuperscript{2} area constraint.
    \end{tablenotes}
\end{threeparttable}
}
\vspace{-0.5em}
\label{tab:arch_config}
\end{table}

%% file: tables/noc_settings.tex
\begin{table}[t]
\caption{NoC Configurations}
\label{tab:noc_settings}
\centering
\scalebox{0.75}{
\begin{threeparttable}
\setlength{\tabcolsep}{1.5pt}
\begin{tabular}{rcccccc}
\hline
                                                             & \textbf{\begin{tabular}[c]{@{}c@{}}Level 3 \\ Topo\end{tabular}} & \textbf{\begin{tabular}[c]{@{}c@{}}Level3 BW\\ (GB/s)\end{tabular}} & \textbf{\begin{tabular}[c]{@{}c@{}}Level 2 \\ Topo\end{tabular}} & \textbf{\begin{tabular}[c]{@{}c@{}}Level2 BW\\ (GB/s)\end{tabular}} & \textbf{\begin{tabular}[c]{@{}c@{}}Level 1\\ Topo\end{tabular}} & \textbf{\begin{tabular}[c]{@{}c@{}}Level1 BW\\ (GB/s)\end{tabular}} \\ \hline
\textbf{\begin{tabular}[c]{@{}r@{}}H100*\\ 32x\end{tabular}} & \begin{tabular}[c]{@{}c@{}}SWITCH\\ 4\end{tabular}               & 200                                                                 & \begin{tabular}[c]{@{}c@{}}SWITCH\\ 8\end{tabular}               & 450                                                                 & -                                                               & -                                                                   \\
\textbf{TMS1}                                                & \begin{tabular}[c]{@{}c@{}}TORUS\\ 4x4\end{tabular}              & 75                                                                  & \begin{tabular}[c]{@{}c@{}}MESH\\ 2x2\end{tabular}               & 384                                                                 & \begin{tabular}[c]{@{}c@{}}SWITCH\\ 4\end{tabular}              & 800                                                                 \\
\textbf{TMS2}                                                & \begin{tabular}[c]{@{}c@{}}TORUS\\ 4x4\end{tabular}              & 75                                                                  & \begin{tabular}[c]{@{}c@{}}MESH\\ 2x2\end{tabular}               & 768                                                                 & \begin{tabular}[c]{@{}c@{}}SWITCH\\ 4\end{tabular}              & 800                                                                 \\
\textbf{TMS3}                                                & \begin{tabular}[c]{@{}c@{}}TORUS\\ 2x2\end{tabular}              & 112                                                                 & \begin{tabular}[c]{@{}c@{}}MESH\\ 4x4\end{tabular}               & 576                                                                 & \begin{tabular}[c]{@{}c@{}}SWITCH\\ 4\end{tabular}              & 1200                                                                \\
\textbf{TMS4}                                                & \begin{tabular}[c]{@{}c@{}}TORUS\\ 2x2\end{tabular}              & 60                                                                  & \begin{tabular}[c]{@{}c@{}}MESH\\ 4x4\end{tabular}               & 307                                                                 & \begin{tabular}[c]{@{}c@{}}SWITCH\\ 4\end{tabular}              & 640                                                                 \\
\textbf{TMS5}                                                & \begin{tabular}[c]{@{}c@{}}TORUS\\ 2x2\end{tabular}              & 75                                                                  & \begin{tabular}[c]{@{}c@{}}MESH\\ 2x2\end{tabular}               & 384                                                                 & \begin{tabular}[c]{@{}c@{}}SWITCH\\ 16\end{tabular}             & 600                                                                 \\
\textbf{TMS6}                                                & \begin{tabular}[c]{@{}c@{}}TORUS\\ 4x4\end{tabular}              & 100                                                                 & \begin{tabular}[c]{@{}c@{}}MESH\\ 2x2\end{tabular}               & 384                                                                 & \begin{tabular}[c]{@{}c@{}}SWITCH\\ 4\end{tabular}              & 800                                                                 \\
\textbf{TMS7}                                                & \begin{tabular}[c]{@{}c@{}}TORUS\\ 4x4\end{tabular}              & 75                                                                  & \begin{tabular}[c]{@{}c@{}}MESH\\ 2x2\end{tabular}               & 384                                                                 & \begin{tabular}[c]{@{}c@{}}SWITCH\\ 4\end{tabular}              & 960                                                                 \\
\textbf{TMM8}                                                & \begin{tabular}[c]{@{}c@{}}TORUS\\ 2x2\end{tabular}              & 75                                                                  & \begin{tabular}[c]{@{}c@{}}MESH\\ 4x4\end{tabular}               & 384                                                                 & \begin{tabular}[c]{@{}c@{}}MESH\\ 2x2\end{tabular}              & 800                                                                 \\ \hline
\end{tabular}
\begin{tablenotes}
        \footnotesize
        \item *In this paper, we use \textbf{T} for TORUS, \textbf{M} for MESH, and \textbf{S} for SWITCH, and arrange them according to an L3–L2–L1 hierarchy.
        \item * The H100*32× topology is expressed using an $800\text{mm}^2$ die as the minimum unit, whereas all other topologies use a $100\text{mm}^2$ 3D-stack cluster as the basic building block.
    \end{tablenotes}

\end{threeparttable}
}
\vspace{-0.5em}
\end{table}

%% file: tex/4_evaluation_v3.tex
\begin{figure}[t]
    \centering
    \includegraphics[width=0.95\linewidth]{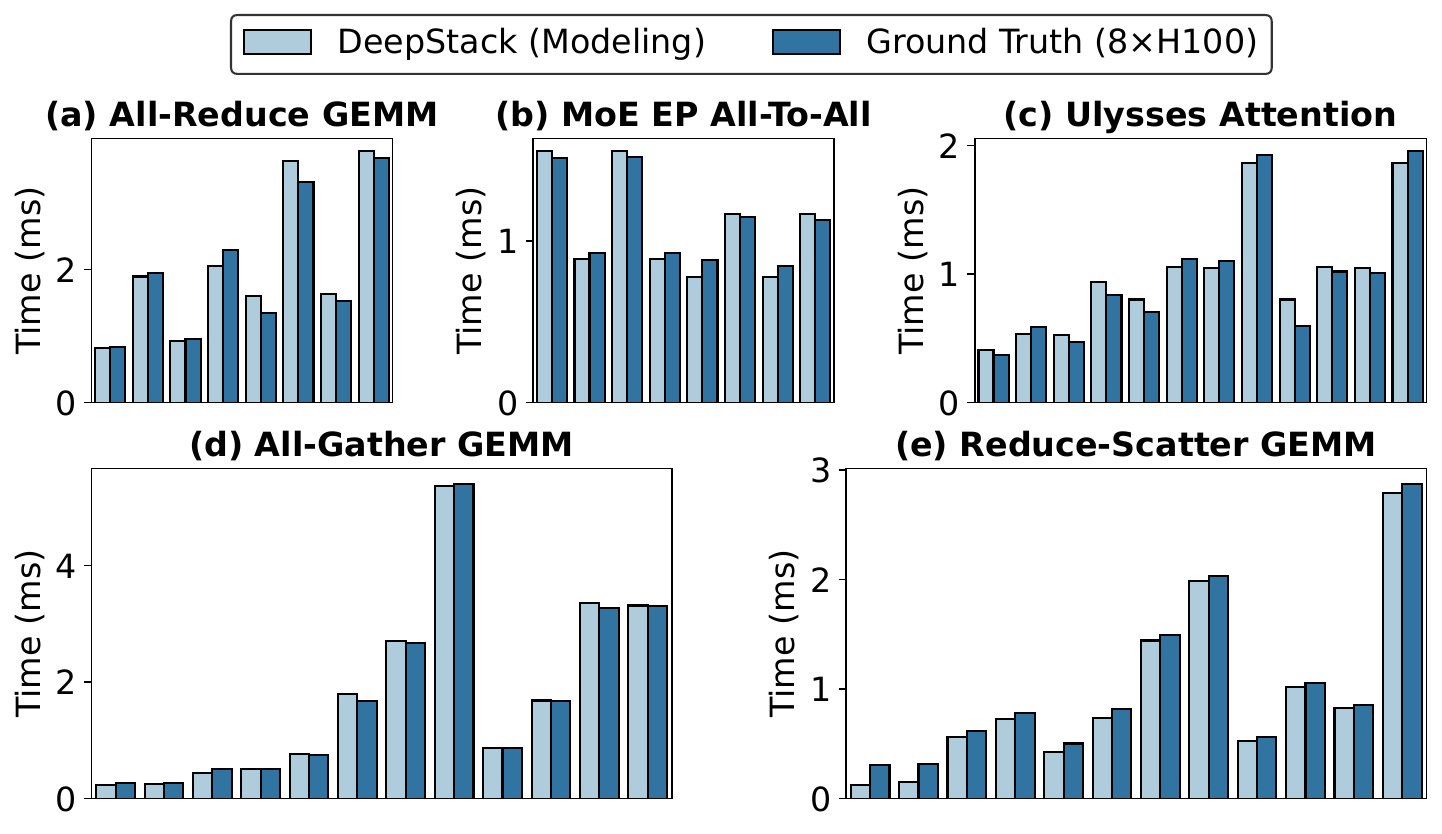}
    \vspace{-0.5em}
    \caption{
    \oursys{} modeling accuracy compared to 8$\times$H100 GPUs. The x-axis enumerates Triton-distributed kernels with different shapes.
    }
    \vspace{-1em}
    \label{fig:modeling_ref_h100}
\end{figure}

\section{Evaluation and Design Implications}\label{sec:eval}

\subsection{Experimental Setup}

\textbf{Workloads \& Metrics.}
We evaluate \oursys{} using Llama-3.3-70B/405B~\cite{dubey2024llama3} (A16W16, A/W denote activation/weight), DeepSeek-R1/V3~\cite{guo2025deepseek-r1,liu2024deepseek-v3} (A16W8), and Qwen3-235B~\cite{yang2025qwen3} (A16W16), with batch sizes ranging from 1 to 1024. Sequence lengths follow~\cite{inferencemax_semianalysis}, with both input and output lengths set to 1024.
Following~\cite{nvidia2025bandwidth_compute_sync_capacity_all_you_need}, we report TTFT for prefill, and UTPS/STPS for both prefill and decoding.

\textbf{Platforms.}
Real-hardware experiments are conducted on H100-SXM~\cite{nvidia2023h100} and B200~\cite{nvidia2024blackwell} systems equipped with AMD EPYC or Intel Xeon hosts. We use Triton-Distributed~\cite{zheng2025tritondistributed} for multi-GPU kernel implementation, vLLM v0.15.1~\cite{kwon2023vllm} with CUDA 13.0 for multi-GPU LLM inference, and ASTRA-Sim v2.0~\cite{astra-simv2} for baseline NoC simulation.

\begin{figure}[t]
    \centering
    \includegraphics[width=0.95\linewidth]{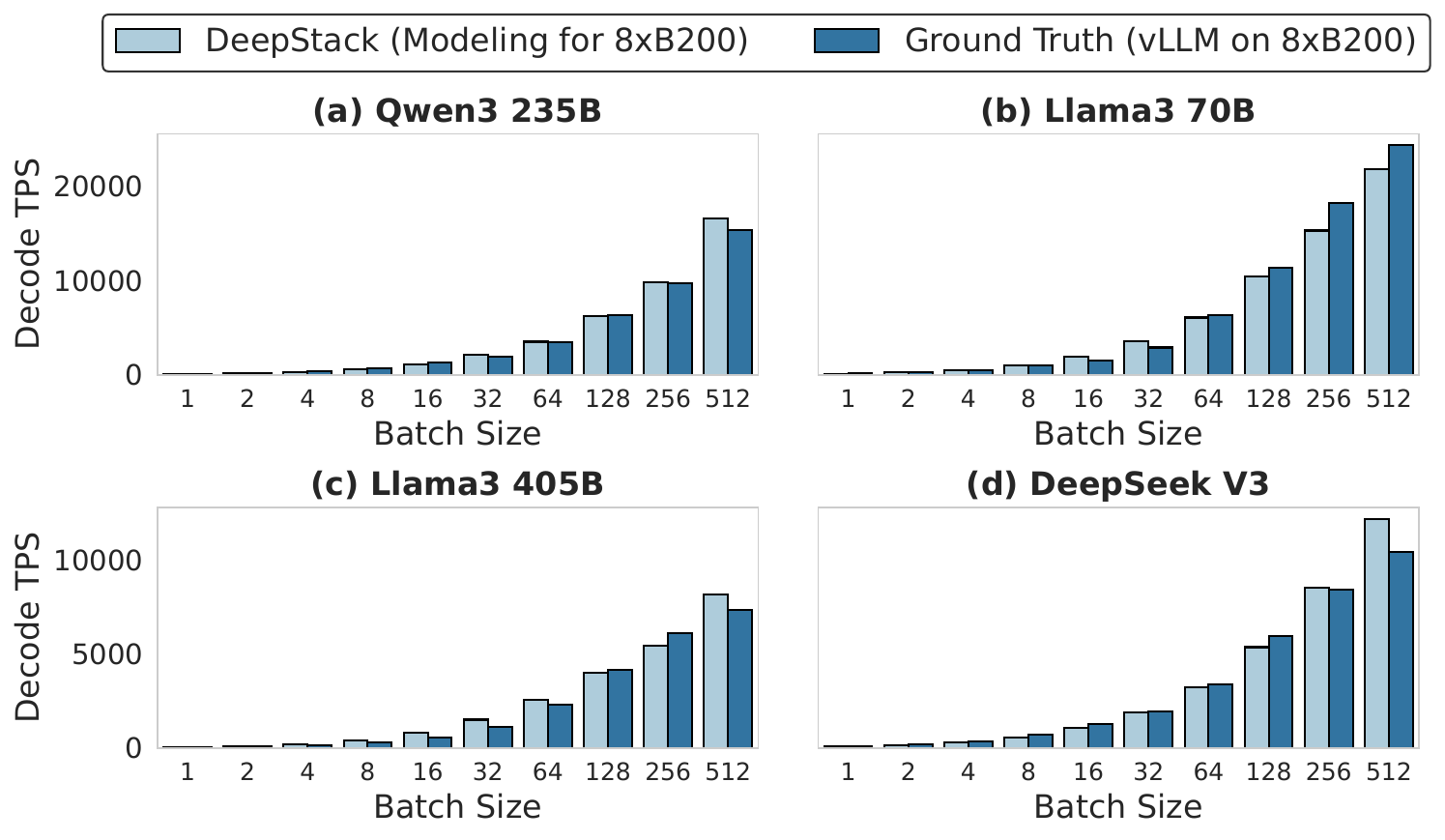}
    \vspace{-0.5em}
    \caption{\oursys{} modeling accuracy for vLLM decode throughput on an
    8$\times$B200 GPU system across four models and ten batch sizes.}
    \vspace{-0.5em}
    \label{fig:modeling_ref_b200}
\end{figure}

\begin{figure}[t]
    \includegraphics[width=\linewidth]{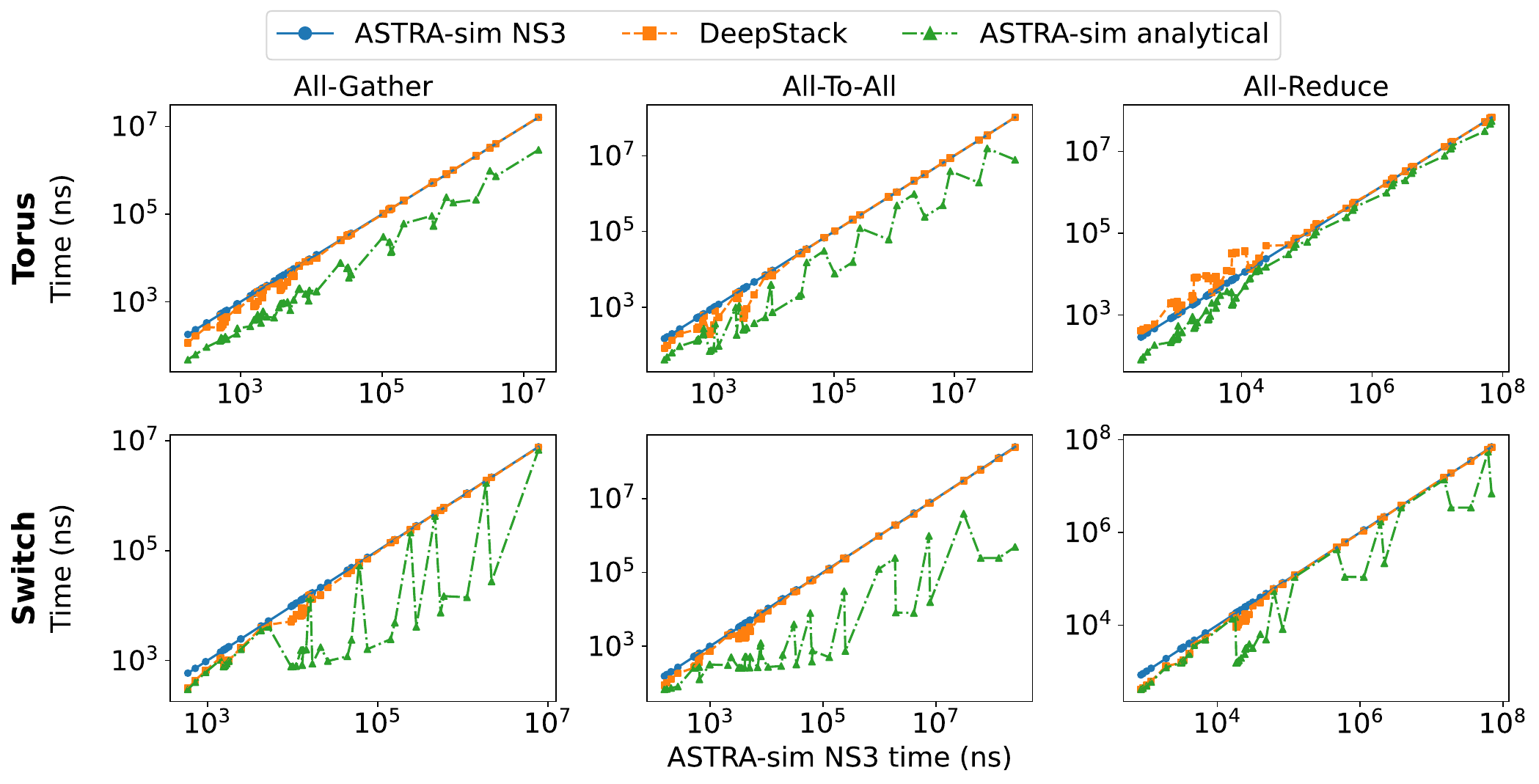}
    \vspace{-1em}
    \caption{
        \oursys{} modeling accuracy vs.\ ASTRA-sim (NS-3 and analytical backends) across Switch and Torus topologies.
        \oursys{} matches NS-3 fidelity with up to $\sim10^5\times$ speedup.
    }
    \vspace{-1em}
    \label{fig:noc_modeling_accuracy}
\end{figure}

\begin{figure*}
\centering
\includegraphics[width=0.9\linewidth]{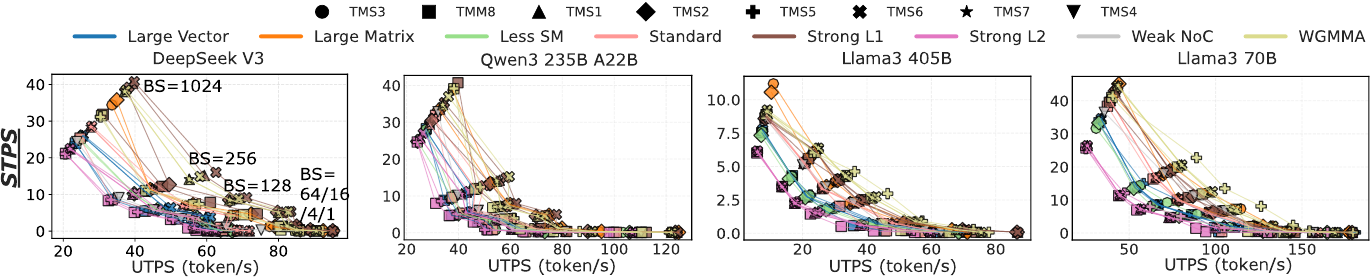}
\vspace{-0.5em}
\caption{
Pareto frontiers from \oursys{}'s DSE across design points. A representative subset is shown for clarity.
Each point on a curve is one batch size (\mbox{BS\,=\,1--1024}) with the parallelism strategy re-searched per point.
}
\vspace{-0.5em}
\label{fig:arch_dse}
\end{figure*}

\begin{figure*}[t]
    \centering
    \includegraphics[width=\linewidth]{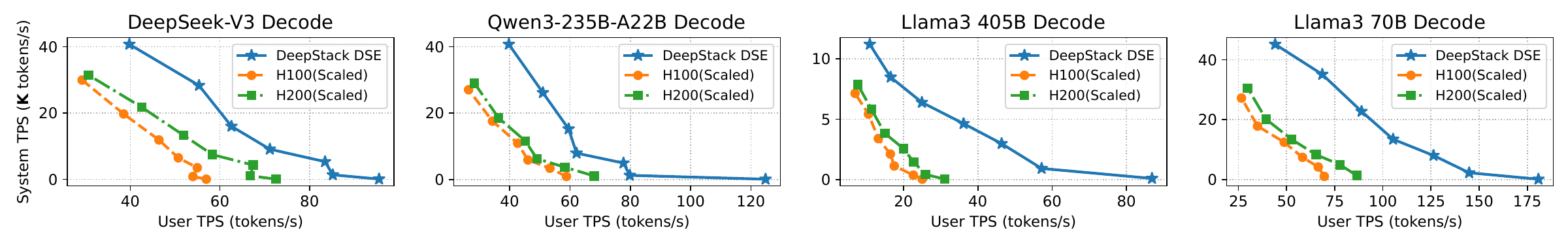}
    \vspace{-2em}
    \caption{
    UTPS/STPS decoding performance comparison.
    \oursys{} DSE denotes the best configuration searched by our framework.
    3D-stacked architectures deliver substantial gains in the memory-bound decode phase.
    }
    \vspace{-0.5em}
    \label{fig:3d_2d_compare_stps_prefill}
\end{figure*}

\textbf{3D Memory Configuration.} By default, the system adopts four 3D-stacked DRAM layers with direct connectivity to the compute die. \S~\ref{ssec:case_study} further evaluates the impact of varying the number of DRAM layers and memory connectivity.

\textbf{Compute Configuration.} All designs are normalized to 7\,nm under a fixed $25{,}600\,\text{mm}^2$ total area budget.
Baselines (Table~\ref{tab:arch_config}) are H100*/H200* scaled to 7\,nm using $800\,\text{mm}^2$ units, with a $1.5\times$ down-scaling factor applied to the 4\,nm H100 (reducing throughput to 504.6\,TFLOPS) based on transistor density differences~\cite{nvidia2020ampere,nvidia2023h100}.

\textbf{NoC Configuration.} To evaluate the impact of interconnect design, we explore a range of NoC configurations that vary in bandwidth and topology, as summarized in Table~\ref{tab:noc_settings}.

\subsection{Modeling Accuracy}\label{subsec:model_acc}

\subsubsection{\textbf{End-to-End LLM Serving Modeling Accuracy}}
\label{sssec:h100_ref}

We cross-validate our model against in-house 3D designs using Cadence Palladium~\cite{cadence_palladium} cycle-accurate emulation (e.g., M3584$\times$N3584$\times$K4096), which achieves less than 5\% error relative to the reference. Because commercial 3D-stacked DRAM systems remain limited, we additionally validate against 8$\times$H100 and 8$\times$B200 NVLink systems to further demonstrate the accuracy of our compute and NoC modeling.
On H100 (Fig.~\ref{fig:modeling_ref_h100}) using Triton-Distributed kernels from LLM Attention and MoE blocks, \oursys{} achieves 3.97\% average error for All-Gather GEMM and 8.4\% weighted error across all kernels.
On B200 (Fig.~\ref{fig:modeling_ref_b200}), end-to-end validation with vLLM~\cite{kwon2023vllm} and CUDA Graphs yields a 12.92\% pointwise MAPE over the 40 plotted model--batch-size points. Residual gaps stem from implementation details (e.g., FlashMLA dynamic KV-splitting) outside our analytical scope.

\subsubsection{\textbf{Comparison with State-of-the-Art Simulators}}
\label{sssec:astrasim_ref}
To compare with prior work, we validate \oursys{} against ASTRA-sim-v2 (NS-3 backend~\cite{riley2010ns-3}) across bandwidths (32--512\,GB/s), latencies (32--512\,ns), topologies (Mesh, Torus, Switch), and collective algorithms.
Network buffers are sufficiently provisioned in all NS-3 experiments.
As shown in Fig.~\ref{fig:noc_modeling_accuracy}, \oursys{} achieves weighted errors of only 2.12\% (Switch) and 1.62\% (Torus) versus the NS-3 backend, while their analytical baseline deviates by up to 58\%.
Crucially, \oursys{} delivers up to \textbf{100,000$\times$ speedup} (0.1\,s vs.\ 3\,h for GiB-scale collectives), enabling billion-scale DSE at discrete-event-level accuracy.

\subsubsection{\textbf{Thermal Model Cross-Check}}
\label{sssec:thermal_ref}
The compact model is a linear fit to sampled ANSYS Fluent points~(\secref{subsec:power_model}) and is cross-checked against HotSpot~\cite{huang2006hotspot,han2021_hotspot7} on a documented 3D stack: a \mbox{10$\times$10\,mm} tile, \mbox{150\,$\mu$m} logic die, \mbox{50\,$\mu$m} per DRAM layer, and an equivalent inter-layer BEOL/low-k bonding interface. For the conservative all-logic-layer power distribution, HotSpot yields \mbox{$R(m){=}0.561{+}0.0098m$}, matching our model within \mbox{${\sim}2\%$} across \mbox{$m\in[1,12]$}; other power distributions reduce junction temperature by up to \mbox{1.8\,\textdegree C}. To be safe, we adopt the conservative estimate.

%% file: tex/5_case_study_v4.tex
\subsection{System-Hardware Co-Design Exploration}\label{subsec:eval_codesign}
\label{sssec:eval_dse}

The fast and accurate modeling (\S\ref{subsec:model_acc}) enables efficient design space exploration across both hardware and system.
Fig.~\ref{fig:arch_dse} presents the performance (STPS \& UTPS) under different typical settings (Tables~\ref{tab:arch_config} and~\ref{tab:noc_settings}) and workloads (different models \& batch sizes), showing only the best-performing designs under each setting.

\underline{\textbf{Key Takeaway 1:}} For compute-bound prefill, \texttt{Large-Matrix} configurations dominate the Pareto frontier with overlapping NoC variants. In contrast, small-batch decode is NoC-sensitive, with bandwidth and hierarchy differences causing large STPS variations.
Model-specific NoC preferences emerge: Llama favors \texttt{TMS5} (stronger L1 switching) due to its reliance on TP, while DeepSeek benefits from \texttt{TMS6} (stronger L3/off-chip bandwidth) driven by large-scale EP.

\underline{\textbf{Key Takeaway 2:}} The optimal parallelism strategy shifts dramatically with batch size: TP dominates at small BS, but drops monotonically as BS grows and communication overhead becomes the bottleneck. PP shows the opposite trend, as larger batches amortize the pipeline bubble. For MoE models, EP is increasingly favored at large BS.

\subsection{Quantitative Analysis of 3D vs.\ 2.5D Systems }\label{subsec:3dvs2d}
To quantitatively analyze the benefits of 3D-stacked DRAM for distributed LLM inference, we compare 3D designs against H100* / H200* 2.5D baselines (Table~\ref{tab:arch_config}) scaled to 7 \, nm, modeling a 32-GPU cluster with NVSwitch intra-node and InfiniBand inter-node~\cite{nvidia_dgx_h100_guide}. For \oursys{}, we perform a comprehensive hardware--software co-search across compute configurations and NoC topologies (Table~\ref{tab:noc_settings}).
All candidates, including all 2.5D and 3D designs, are constrained to an identical total area budget of $25{,}600\,\text{mm}^2$.

Fig.~\ref{fig:3d_2d_compare_stps_prefill} shows the Pareto frontiers derived from over one million design points.
In the memory-bound decoding, 3D-stacked designs significantly outperform 2.5D baselines, with speedups of \textbf{1.30--1.48$\times$} at BS\,=\,1024 and up to \textbf{2.79$\times$} at BS\,=\,1, where per-token memory traffic dominates.
These results use a fixed \textbf{4-layer} stacking. DRAM layer and NoC co-optimization (\S\ref{ssec:case_study}) further pushes the DeepSeek-V3 advantage to \textbf{1.73$\times$} at BS\,=\,1024 (Table~\ref{tab:ablation}).
In the compute-bound prefill phase, 3D designs perform comparably to H200*, as the latter's higher peak TFLOPs and larger capacity (141\,GiB HBM3e vs.\ 64\,GiB 3D-DRAM) offset the bandwidth advantage of 3D designs.

\underline{\textbf{Key Takeaway 3:}}
The 3D-stacked design serves as an effective accelerator for the decode phase, achieving up to 2.79$\times$ STPS improvement. During compute-bound prefill, its performance remains comparable to that of 2.5D baselines.

\subsection{Case Study: 3D DRAM Layer Design Space}
\label{ssec:case_study}

Central design questions in 3D-stacked architectures are: \emph{i) how many DRAM layers should be stacked, and ii) how many should be actively connected via hybrid bonding?} Stacked layers contribute capacity and increase thermal resistance; connected layers contribute bandwidth but consume additional power and PHY/controller area.
We explore this design space and derive quantitative guidance.

\subsubsection{\textbf{Layer--Bandwidth Trade-off}}

\begin{figure}
\centering
\includegraphics[width=1.0\linewidth]{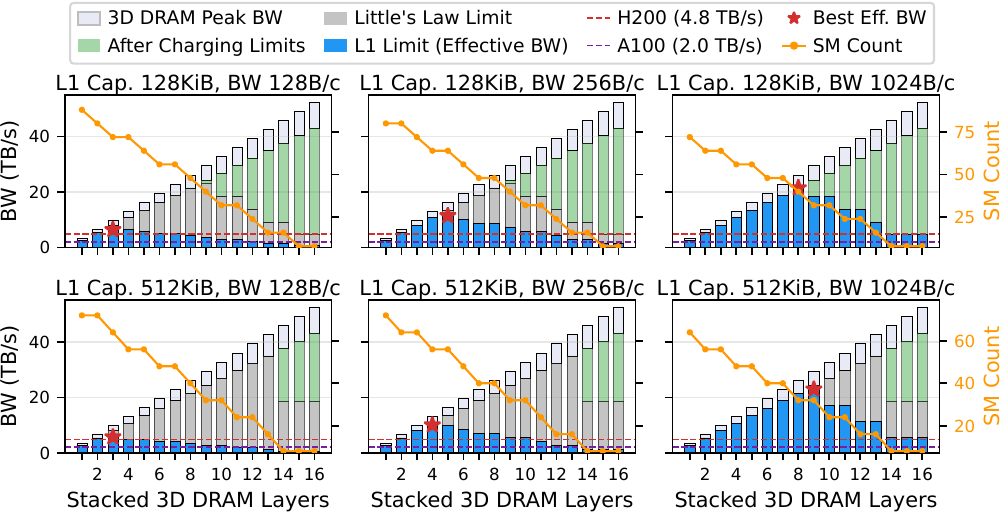}
\vspace{-1em}
\caption{Theoretical and effective DRAM bandwidth breakdown \& SM count as a function of stacked 3D DRAM layers (stack\,=\,connected, 800\,mm$^2$, 7\,nm
process
). A representative subset is shown for clarity.
}
\vspace{-0.5em}
\label{fig:bw_vs_stacked_layers}
\end{figure}

We sweep L1 capacity (128/\hspace{0pt}256/\hspace{0pt}512\,\hspace{0pt}KiB) and bandwidth (128/256/512/1024\,B/cycle). Fig.~\ref{fig:bw_vs_stacked_layers} shows a representative subset.
Counter-intuitively, \textbf{effective bandwidth peaks at fewer than 10 layers under 7\,nm}. Beyond this point, Little's law and L1 bandwidth constraints cause effective bandwidth to decline even as theoretical bandwidth rises. We therefore prune configurations exceeding 9 layers from the subsequent design space.

\underline{\textbf{Key Takeaway 4:}} {DRAM layer stacking exhibits an inverted-U curve: beyond ${\sim}$9 layers, effective bandwidth \emph{decreases} despite increasing theoretical bandwidth, mainly due to buffering limitation.}
For instance, a 12-layer stack delivers 39.6\% \emph{lower} effective bandwidth than the 9-layer peak, so over-stacking could hurt.

\subsubsection{\textbf{End-to-End Performance vs.\ Stacked Layers}}

\begin{figure}
\centering
\includegraphics[width=0.98\linewidth]{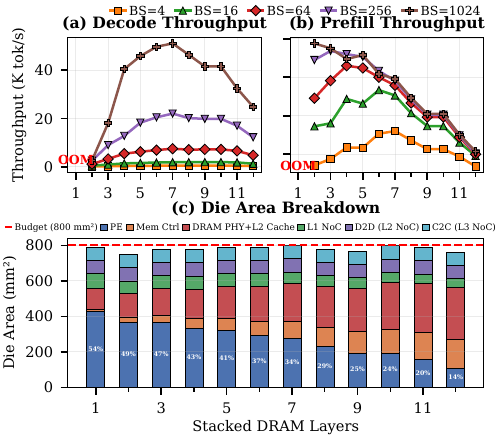}
\vspace{-0.5em}
\caption{
End-to-end TPS for DeepSeek-V3 and area breakdown across different DRAM stacking layers.
The optimal layer depends on batch size and inference phase.
}
\vspace{-0.5em}
\label{fig:e2e_perf_vs_stacked_layers}
\end{figure}

Fig.~\ref{fig:e2e_perf_vs_stacked_layers} shows the end-to-end throughput and area breakdown, sweeping stacked layers (all connected) under a fixed area budget where DRAM layers compete directly with compute SMs for die area.
The optimal stacking depth varies significantly with workload:

\begin{itemize}[leftmargin=*]
    \item \textbf{Small-batch decode (BS=4):} favors deep stacks (${\sim}$9 layers) to maximize bandwidth for memory-bound single-token generation.
    \item \textbf{Large-batch decode (BS=1024):} the optimum drops to 6--7 layers, as the workload becomes partially compute-bound and the area consumed by additional DRAM layers is better allocated to SMs.
    \item \textbf{Small-batch prefill:} prefers ${\sim}$7 layers, balancing bandwidth with compute capacity.
    \item \textbf{Large-batch prefill:} only 2 layers are optimal, as the workload is fully compute-bound and every additional DRAM layer displaces productive compute area.
\end{itemize}

\underline{\textbf{Key Takeaway 5:}} This analysis reveals a design insight beyond conventional prefill--decode (PD) disaggregation: \emph{batch size defines a more fundamental architectural divide than the prefill/decode distinction}. The optimal hardware maps to three architecture pools: (1)~large-batch prefill uses shallow/no stacks to maximize compute and avoid stacking cost/thermal risk; (2)~small-batch prefill and large-batch decode share a moderate-depth, balanced design; and (3)~small-batch decode uses deep stacks for the highest-UTPS, bandwidth-bound tier. Thus, batch-size-aware hardware disaggregation may be more effective than PD disaggregation alone.

\subsubsection{\textbf{Connected vs.\ Stacked Layers: Decoupled Design}}
\label{ssec:connected_vs_stacked}

\begin{figure*}
\centering
\includegraphics[width=1.0\linewidth]{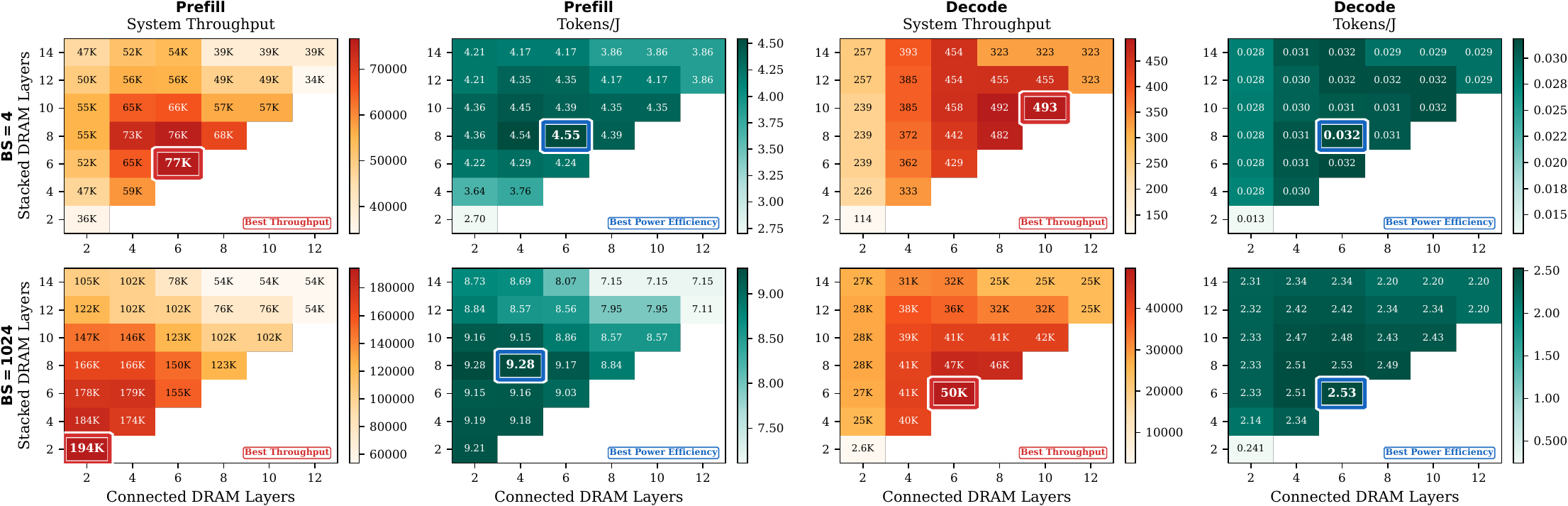}
\vspace{-1em}
\caption{DSE heatmaps for throughput-optimal and energy-optimal designs across stacked ($m$) and connected ($n$) DRAM layer configurations. Energy-efficient designs consistently favor more stacked but fewer connected layers.}
\vspace{-1em}
\label{fig:e2e_perf_stacked_vs_connected_layers}
\end{figure*}

We differentiate stacked layers ($m$, capacity) from connected layers ($n \leq m$, bandwidth). Unconnected layers add capacity without the power and PHY area overhead of active connections but increase thermal resistance.
As shown in Fig.~\ref{fig:e2e_perf_stacked_vs_connected_layers}, the throughput-optimal and energy-optimal designs diverge significantly:

\textbf{Throughput-optimal configurations} maximize connected layers to saturate bandwidth, with optimal $(m,n)$ shifting from $(6,6)$ at BS=4 to $(2,2)$ at BS=1024 for prefill. This $\Delta m{=}4$ gap is driven by batch size, whereas the prefill-vs-decode gap at the same BS can be smaller or even zero, further confirming that batch size drives architectural divergence more than inference phase.

\underline{\textbf{Key Takeaway 6:}} {Energy-efficient and throughput-optimal designs require fundamentally different architectures.}
\textbf{Energy-efficient configurations} consistently favor more stacked but fewer connected (idle) layers, operating at
$0.5$--$0.7$\,W/mm$^2$
({\(\sim\)}6--48\% below throughput-optimal) with 3--24\% tokens/J gains. These designs combine deep capacity stacking with relatively few connected layers and larger on-chip buffers, improving data reuse and reducing DRAM accesses while admitting communication-friendlier parallelism with lower TP. In the evaluated MoE cases, this shift is EP-concentrated (e.g., tp1/ep32 vs.\ tp4/ep4/dp4/pp4), partitioning MoE weights across devices to reduce memory-access energy.

\subsubsection{\textbf{Thermal Feasibility}}

\begin{figure}
\centering
\includegraphics[width=1.0\linewidth]{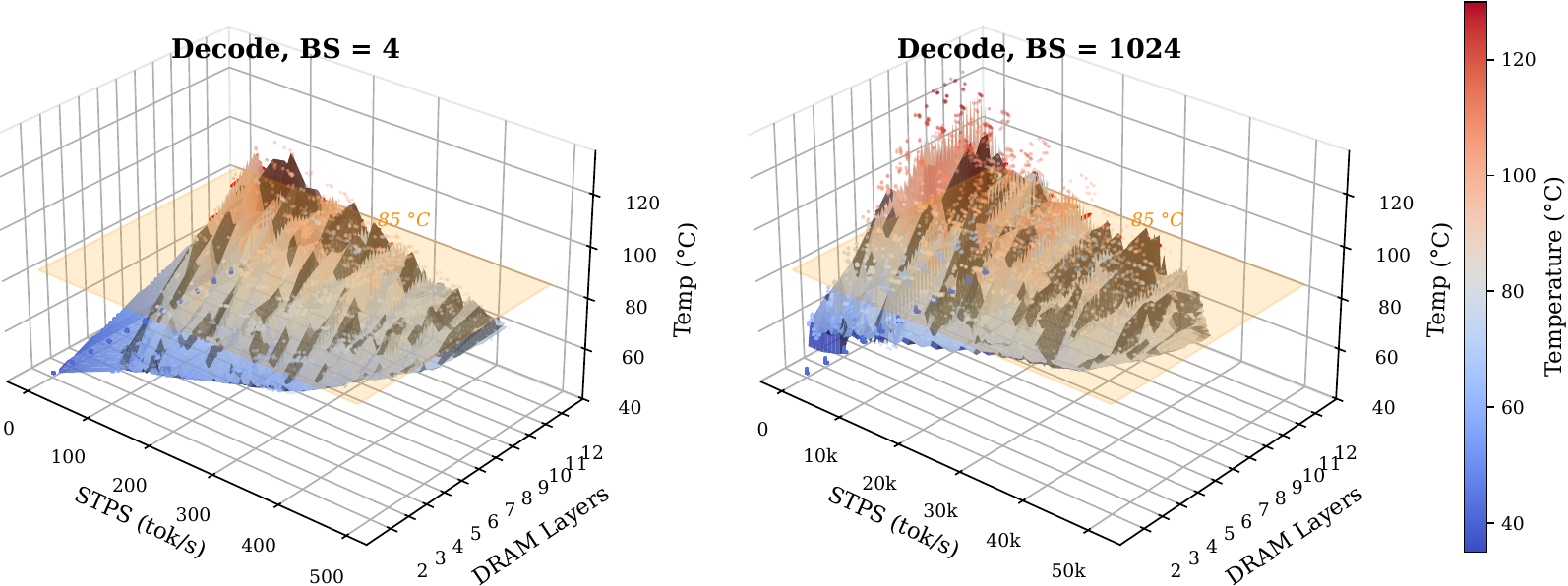}
\caption{
Temperature distribution across the 3D DRAM design space. Large-batch decode is the most thermally challenging regime. High-throughput configurations within the safe 85$^\circ$C envelope exist.
}
\label{fig:temp_vs_stacked_layers}
\end{figure}

\begin{figure}
\centering
\includegraphics[width=1.0\linewidth]{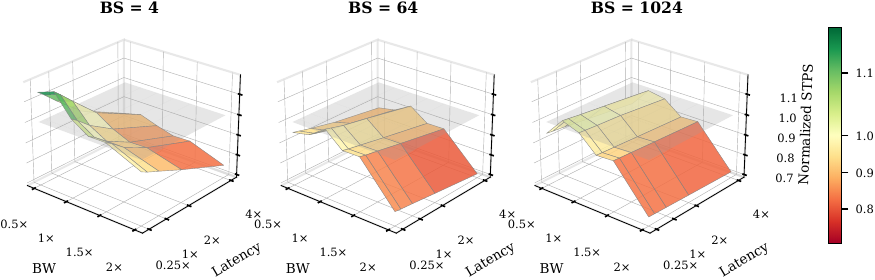}
\caption{
NoC bandwidth and hop latency sensitivity. Reducing bandwidth frees die area for additional SMs.
}
\vspace{-0.5em}
\label{fig:noc_latency_bandwidth_sensitivity}
\end{figure}

We conduct thermal analysis for different workloads.
As shown in Fig.~\ref{fig:temp_vs_stacked_layers}, large-batch decode is the
most thermally challenging regime, with many configurations exceeding
85$^\circ$C. Thermal violations are less frequent for prefill and small-batch
decode. The search nevertheless finds thermally feasible configurations with
high throughput in every evaluated regime. Larger on-chip
buffers and coarser tiling reduce DRAM access frequency, lowering power without
sacrificing throughput.
Notably, all thermally feasible designs identified by our DSE operate below ${\sim}$0.8\,W/mm$^2$, well under the ${\geq}$1.34\,W/mm$^2$ already sustained by the NVIDIA Vera Rubin GPU~\cite{semianalysis_verarubin_power} with production liquid-cooling solutions.

\underline{\textbf{Key Takeaway 7:}} {Thermal limits do not rule out high-throughput designs, but they constrain the viable design space, especially for large-batch decode. Thermal-aware DSE is therefore necessary to identify designs that are both high-performance and feasible.}

\subsubsection{\textbf{NoC Sensitivity Analysis}}

As shown in Fig.~\ref{fig:noc_latency_bandwidth_sensitivity}, we perform a sensitivity study on NoC bandwidth and hop latency, scaling all three hierarchical layers defined in Table~\ref{tab:noc_settings} simultaneously. Importantly, we co-optimize NoC bandwidth with compute area: reducing bandwidth frees die area for additional SMs (e.g., $0.75\times$ BW yields 9 SMs vs.\ 8 at $1\times$ BW), while increasing to $2\times$ BW leaves only 5 SMs.

At BS=1024, latency has minimal impact (3.6\% STPS variation over a $0.25\times$-to-$4\times$ sweep), as abundant parallelism hides communication latency. Bandwidth matters only through its area trade-off: $0.75\times$ BW gains 2.4\% from one extra SM, whereas $2\times$ BW loses 27.2\% from fewer SMs.
At BS=4, the trend reverses: latency becomes dominant (up to 31.3\% variation, $1.36\times$ speedup at $0.25\times$), and reducing BW to $0.75\times$ still improves STPS by 8.8\% by freeing area for compute.
The overall optimal is $0.75\times$ BW with $0.25\times$ latency, achieving up to 17.3\% improvement for BS=4.

\begin{figure}[t]
\centering
\includegraphics[width=1.0\linewidth]{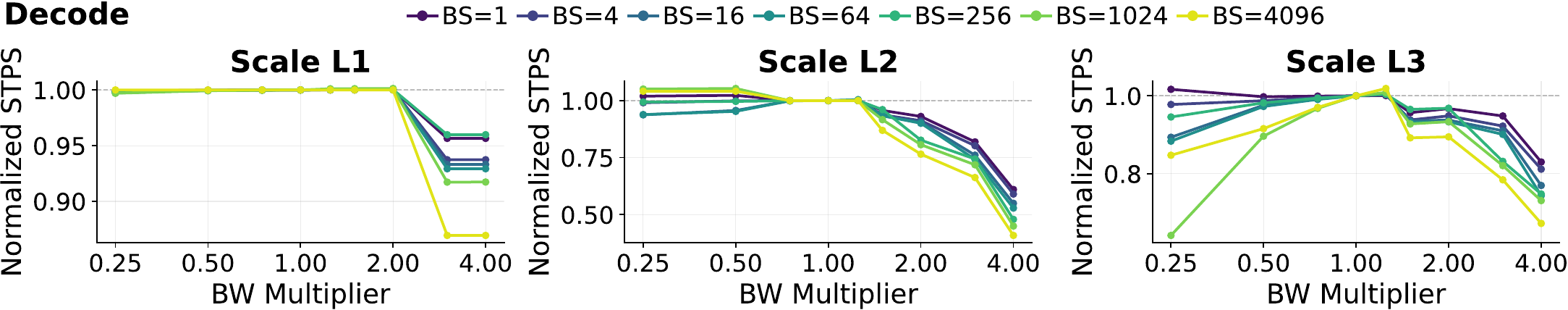}
\includegraphics[width=1.0\linewidth]{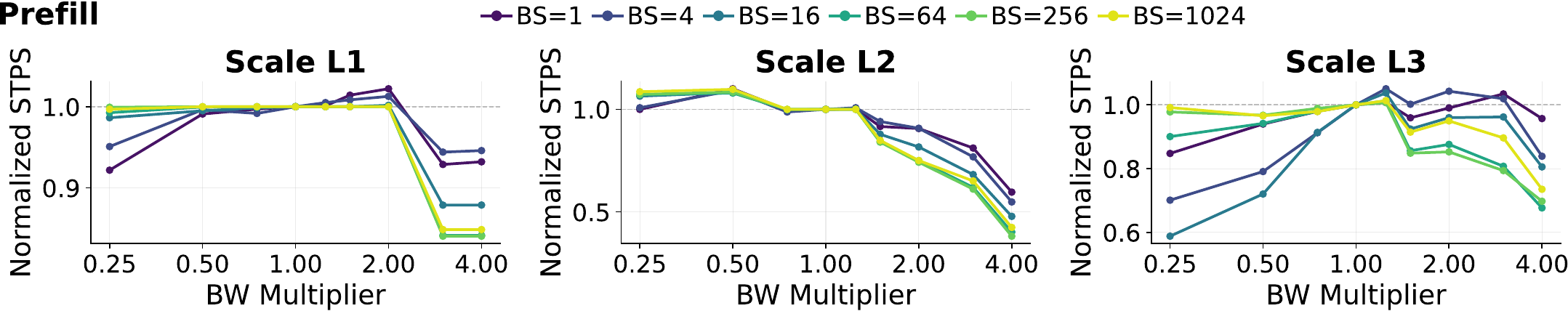}
\caption{Per-layer NoC bandwidth sensitivity. Each hierarchical layer is scaled independently while others remain at baseline. Upper: decode phase; lower: prefill phase.
}
\vspace{-1em}
\label{fig:noc_layer_bw_sensitivity}
\end{figure}

\begin{figure}
    \centering
    \includegraphics[width=1.0\linewidth]{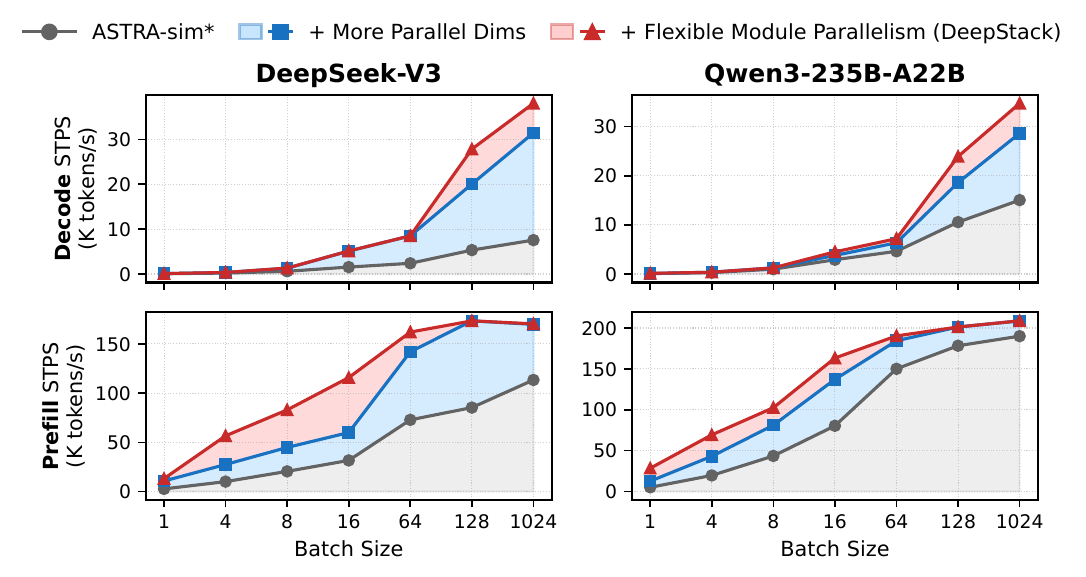}
    \caption{Effectiveness of expanded parallelism in \oursys{}.
    ASTRA-sim* supports only TP/PP/DP.}
    \label{fig:dse-our-ast}
    \vspace{-1em}
\end{figure}

Fig.~\ref{fig:noc_layer_bw_sensitivity} decomposes the contribution of each NoC layer by independently scaling its bandwidth. The sensitivity is non-uniform: L2 (inter-tile) bandwidth can be modestly reduced without performance loss, while L3 (off-chip) benefits from a slight increase, suggesting asymmetric interconnect area allocation across the hierarchy.
This behavior is coupled with the optimal parallelism mapping, re-searched at each scaling point: EP all-to-all needs high bandwidth across its entire scope, DP and PP traffic is light, and no dimension sheds traffic beyond the L2 scope.

\underline{\textbf{Key Takeaway 8:}} NoC provisioning should be batch-aware: small batches are latency-sensitive, while large batches benefit from high L3 bandwidth because the re-searched mapping selects large-scale EP with broad-scope all-to-all traffic. This aligns with NVL72-style high-bandwidth scale-out for large-batch inference.

\begin{table}[t]
\centering
\caption{Ablation study on \oursys{} techniques. All configurations satisfy the area ($32\times800\,\text{mm}^2$) and thermal constraints.}
\label{tab:ablation}
\small
\setlength{\tabcolsep}{4pt}
\scalebox{0.85}{
\begin{tabular}{@{}rlrr@{}}
\toprule
\textbf{\#} & \makecell[l]{\textbf{Ablation Study}} & \textbf{STPS}$_{bs\!=\!4}$ & \textbf{STPS}$_{bs\!=\!1024}$ \\
\midrule
1 & Baseline (ASTRA-sim:DP/TP/PP/FSDP) & 177.1 & 5,729 \\
2 & + Full Parallel(EP/SP/CP/FSDP) & 256.4 (+45\%) & 21,252 (+271\%) \\
3 & + \makecell[l]{Flex. Parallel Across Modules \\ (E.g., EP in MoE and TP in Atten.)} & 256.4 (---) & 24,488 (+15\%) \\
4 & + Search On-chip Arch. & 314.2 (+23\%) & 31,350 (+28\%) \\
5 & + Comm.-Comp. Overlap & 340.5 (+8\%) & 38,061 (+21\%) \\
6 & + Stacking DRAM Layer DSE & 493.3 (+45\%) & 51,095 (+34\%) \\
7 & + NoC DSE & 494.1 (+0.2\%) & 54,280 (+6.2\%) \\
\midrule
  & \textbf{Ablation Gain} & \textbf{2.8$\times$} & \textbf{9.5$\times$} \\
\bottomrule
\end{tabular}}
\vspace{-0.5em}
\end{table}

\subsection{Ablation Study}
\label{ssec:ablation}

\subsubsection{Expanded Parallelism Search Space}
Fig.~\ref{fig:dse-our-ast} isolates the parallelism impact across models at decode BS\,=\,1024. Replacing ASTRA-sim's TP/PP/DP-only space with full parallelism (including EP) and per-module flexibility yields $5.03\times$ for DeepSeek-V3 and $2.31\times$ for Qwen3-235B,
since \oursys{} allows each module to adopt an independent parallelism strategy and automatically inserts the necessary collectives to reconstruct the required input tensors across module boundaries.

\subsubsection{Cumulative Technique Contribution}
Table~\ref{tab:ablation} reports a search-space ablation: a restricted DSE misses up to \mbox{$9.5\times$} modeled throughput, with \mbox{$4.27\times$} from missing parallelism dimensions and \mbox{$2.22\times$} from omitted hardware co-search.
The largest gains come from EP (Step~2) and DRAM-layer DSE (Step~6); Steps~6--7 further raise the equal-area decode advantage over H200* from 1.30$\times$ to 1.73$\times$ at BS=1024.

Notably, omitting even a single parallelism dimension does not merely reduce throughput. It actively misleads the architecture search. At BS=1024, removing EP causes the DSE to converge on a pipeline-parallel-heavy schedule (tp=16, pp=8) and a different chip (8 stacked/connected layers, 5 SMs) that hits the power wall. With EP, the design shifts to an all-to-all regime (ep=32, tp=4, 7 stacked/connected layers, 6 SMs) within thermal limits. At BS=4, the divergence is even starker: stacked layers (7 vs.\ 10), shared memory (128K vs.\ 256K), and SM count (6 vs.\ 4) all differ, producing qualitatively different chips.

\underline{\textbf{Key Takeaway 9:}} {Incomplete schedule search permanently distorts architecture design. The resulting silicon mismatch is irrecoverable by software tuning, underscoring the necessity of co-searching the full system--hardware space before committing to silicon.}

%% file: tex/7_related_v3.tex
\section{Related Work}

\subsection{3D-Stacked DRAM Architecture Exploration}
Recent works explore 3D-stacked or hybrid-bonded DRAM for AI accelerators from different angles.
H2-LLM~\cite{li2025h2} targets near-memory compute on the logic die for low-batch inference. Stratum~\cite{pan2025stratum} exploits heterogeneous access latency in monolithic 3D DRAM for MoE models.
TASA~\cite{he2025tasa} proposes thermal-aware 3D architectures for LLM inference, and AccelStack~\cite{AccelStack} provides a cost modeling framework for hybrid-bonded manufacturing flows.
STCO~\cite{sharda2025system} evaluates 3D-stacked DRAM and V-Cache architectures with area, power, and thermal analysis. Helios~\cite{li2026_3d_Dram_numa_helios} proposes NUMA-centric optimizations for KV-cache placement in 3D DRAM-based LLM serving with TP, PP, and EP. LIMINAL~\cite{nvidia2025bandwidth_compute_sync_capacity_all_you_need} compares SRAM, HBM, and stacked-DRAM architectures at the system level.
These works each address important aspects of 3D design. Our work is complementary in that it provides an end-to-end DSE framework that jointly explores parallelism scheduling, hierarchical NoC design, and thermal constraints in a distributed setting (Table~\ref{tab:tool_comparison_single_safe}).

\subsection{Modeling and DSE Tools}

Single-chip modeling frameworks such as Timeloop~\cite{parashar2019timeloop}, MAESTRO~\cite{kwon2020maestro}, TileFlow~\cite{zheng2023tileflow}, and MIND~\cite{huang2024mind} provide accurate dataflow-level performance estimation but focus on individual accelerator chips. For distributed systems, ASTRA-sim~\cite{astra-simv1,astra-simv2} models hierarchical network topologies and supports TP, PP, DP, and FSDP for training simulation, with its NS-3 backend offering high-fidelity network modeling. LLMCompass~\cite{llmcompass} employs high-level analytical models to rapidly evaluate LLM inference across hardware configurations with TP and PP support. Chiplet Cloud~\cite{peng2023chiplet} explores analytical chiplet-ASIC co-design for LLM serving with an SRAM-centric memory system.
We integrate established analytical primitives, including the \mbox{$\alpha$--$\beta$} link cost, published energy coefficients, and the vendor per-bank bandwidth curve. Our contribution is to extend them to the 3D-stacked domain by capturing fine-grained 3D-DRAM bank characteristics at analytical speed, incorporating a comprehensive inference parallelism space, and providing fast yet accurate distributed network modeling, all unified within a single DSE loop.

\subsection{Thermal Challenges in 3D Stacking}
3D stacking increases power density without enlarging the dissipation area, making thermal a critical constraint.
HotSpot~\cite{Han2022From2T_hotspot7}, CACTI-3DD~\cite{chen2012cacti}, and COMET~\cite{siddhu2022comet} provide compact thermal modeling for 3D-stacked DRAM and multi-core systems.
On the cooling side, emerging solutions include embedded microfluidics~\cite{wang2018microfluidics}, inter-tier liquid cooling~\cite{sridhar20103d}, Through-Chip Microchannels (TCMCs) capable of 14.1\,kW/cm$^2$ dissipation~\cite{ao2024through-chip-cooling}, full immersion cooling~\cite{qiu2017experimental}, and Carbon-Nanotube-based TSVs for enhanced passive conduction~\cite{ha2025overview}.
The NVIDIA Vera Rubin GPU already sustains at least 1.34\,W/mm$^2$ in production with liquid metal TIM and micro-channel cold plates~\cite{semianalysis_verarubin_power}. By integrating thermal modeling into the DSE loop to prune infeasible configurations early, our throughput-optimal 3D-stacked designs stay under 0.8\,W/mm$^2$, comfortably within this envelope.

\subsection{Distributed LLM Inference}

Large-scale LLM training and inference are supported by mature frameworks~\cite{shoeybi2019megatron,narayanan2021efficient-megatron-lm,rasley2020deepspeed,li2023colossal-ai,jax2018jax,lepikhin2020gshard,Pope2022EfficientlyST}. A rich body of work explores parallelism and schedule search~\cite{tofu_wang2019supporting,flexflow_jia2019beyond,zheng2022alpa,zhong2024distserve,zhu2025nanoflow,ko2024dfmodel,lin2024tessel,Huang2018GPipeET,Narayanan2019PipeDreamGP}, extends sequence and expert parallelism for long-context and MoE workloads~\cite{Liu2023RingAW,jacobs2023deepspeed,Wu2024LoongServeES,Hwang2022TutelAM,singh2023hybrid}, optimizes serving runtimes for iteration-level batching, KV-cache management, and prefill--decode disaggregation~\cite{Yu2022OrcaAD,kwon2023vllm,zheng2024sglang,Agrawal2024TamingTT,Sun2024LlumnixDS,Qin2025MooncakeTM}, and fuses collectives with compute to hide network latency~\cite{Jangda2021BreakingTC,Pati2024T3TT,Chang2024FLUXFS,zheng2025tritondistributed}. These efforts target software mapping on \emph{fixed} hardware, and the analytical models they rely on (e.g., Alpa~\cite{zheng2022alpa}, DFModel~\cite{ko2024dfmodel}) use simple linear models that serve as relative cost functions but fall short of accurate performance modeling for emerging 3D-stacked systems with fine-grained 3D characteristics.
Our work complements them by modeling end-to-end inference performance and co-searching hardware architecture with seven parallelism strategies, particularly for emerging 3D-stacked hardware.

%% file: tex/8_conclusion_v4.tex
\section{Conclusion}

We present \oursys{}, a performance modeling and efficient DSE framework for distributed 3D-stacked accelerators. By capturing 3D memory semantics, thermal--power effects, and comprehensive parallelism/scheduling search, \oursys{} enables early-stage exploration of the vast hardware--system co-design space.
Our DSE shows that optimal DRAM stacking depth is governed by a non-trivial bandwidth--area trade-off, batch size creates a more fundamental divide than prefill/decode, and parallelism strategy tightly couples with hardware architecture: optimizing one without the other leads to suboptimal designs irrecoverable by software.
Future work will integrate detailed 3D cost and yield curves into the search objective to enable TCO-aware optimization~\cite{ning2023supply}.
We hope \oursys{} and these findings will guide early-stage architectural decisions for next-generation scalable AI infrastructure.

%% file: tex/artifact_appendix_camera_ready.tex
\appendix[Artifact Appendix]
\begingroup
\setlength{\emergencystretch}{2em}
\newcommand{\artifactpath}[1]{\path{#1}}

\subsection{Abstract}

This artifact provides the implementation and automated evaluation workflow
for DeepStack's analytical modeling and design-space exploration (DSE) of
distributed 3D DRAM-stacked LLM accelerators. It reruns the fixed
configurations underlying Figs.~\ref{fig:modeling_ref_h100}--%
\ref{fig:noc_modeling_accuracy}, Figs.~\ref{fig:arch_dse}--%
\ref{fig:dse-our-ast}, and Table~\ref{tab:ablation}. It checks the results
against bundled, checksummed GPU and simulator references and regenerates
the paper-facing plots. No GPU, model weights, or external simulator is required.

\subsection{Artifact check-list (meta-information)}

{\small
\begin{itemize}
  \item {\bf Algorithm:} analytical latency, throughput, energy, thermal, and
        network modeling with hierarchical DSE.
  \item {\bf Program:} Python~3.11 package and staged \artifactpath{run.sh}
        workflow.
  \item {\bf Compilation/Binary:} no local compilation; four licensed,
        prebuilt CPython~3.11 x86-64 extensions are included.
  \item {\bf Model:} DeepSeek-V3, Qwen3-235B-A22B, Llama-70B, and
        kernel-level workload descriptions; no model weights.
  \item {\bf Data set:} checksummed GPU/vLLM and ASTRA-Sim/NS-3 references,
        fixed-configuration manifests, routing traces, and a recorded
        301{,}728-row decode-DSE population.
  \item {\bf Run-time environment:} x86-64 Linux, GNU Bash, Conda,
        CPython~3.11, and glibc~2.29 or newer.
  \item {\bf Hardware:} 32 CPU cores and 64~GiB RAM recommended; no GPU,
        FPGA, or accelerator.
  \item {\bf Metrics:} weighted MAPE, scaled tokens/s, bandwidth, speedup,
        temperature, and area feasibility.
  \item {\bf Output:} validated CSVs, run/verification manifests, and 16
        figure groups in PNG and PDF.
  \item {\bf Experiments:} 15 fail-fast stages plus independent verification.
  \item {\bf Disk space:} approximately 5~GiB including the environment.
  \item {\bf Preparation time:} approximately 5--10 minutes.
  \item {\bf Experiment time:} 12.2 minutes on the 64-core/128-thread
        packaging host with 32 workers; 30--45 minutes estimated on 32 cores.
  \item {\bf Publicly available?:} Yes, via the public GitHub repository.
  \item {\bf Licenses:} Apache-2.0 for source and data; bundled extensions use
        LicenseRef-DeepStack-AE-Binary-1.0.
  \item {\bf Automation:} repository-local scripts; no external framework.
  \item {\bf Artifact:}
        \mbox{\href{https://github.com/tile-ai/DeepStack/tree/ae}{%
          \nolinkurl{https://github.com/tile-ai/DeepStack/tree/ae}}}\\
        Zenodo:
        \mbox{\href{https://zenodo.org/records/21351389}{%
          \nolinkurl{https://zenodo.org/records/21351389}}}.
\end{itemize}
}

\subsection{Description}

\subsubsection{How to access}

Public artifact branch:
\href{https://github.com/tile-ai/DeepStack/tree/ae}{%
  \nolinkurl{https://github.com/tile-ai/DeepStack/tree/ae}}.
Clone it with:
\begin{lstlisting}[language=bash,aboveskip=1pt,belowskip=1pt]
git clone --branch ae \
  https://github.com/tile-ai/DeepStack.git && cd DeepStack
\end{lstlisting}
Then follow \artifactpath{README.md}.

\subsubsection{Hardware dependencies}

Use a 64-bit x86 Linux host with 32 cores, 64~GiB RAM, and 5~GiB free
storage. Lower core counts work with fewer workers and longer runtime. GPU and
simulator measurements are bundled reference inputs and are not rerun.

\subsubsection{Software dependencies}

The artifact requires GNU Bash, Conda, CPython~3.11, and glibc~2.29 or newer.
\artifactpath{./setup.sh} installs pinned dependencies. Four licensed prebuilt
extensions provide protected model components; no compiler, CUDA, external
simulator, or proprietary EDA tool is needed.

\subsubsection{Data sets and models}

All required references, configurations, routing traces, and plotting data are
included and covered by \artifactpath{SHA256SUMS}. The artifact contains
workload descriptions and routing metadata, but no prompts, weights, or hidden
states. The complete area model and die-area breakdown are not distributed.
Reference feasibility exposes only a feasible-SM count from a finite compiled
capacity table.

\subsection{Installation}

\begin{lstlisting}[language=bash]
./setup.sh
conda run --no-capture-output -n deepstack-ae \
  ./run.sh doctor
\end{lstlisting}

Installation succeeds when the second command prints \code{doctor: PASS}.

\subsection{Experiment workflow}

\begin{lstlisting}[language=bash]
conda run --no-capture-output -n deepstack-ae \
  ./run.sh quick --model-version both
conda run --no-capture-output -n deepstack-ae \
  ./run.sh reproduce --workers 32 \
  --model-version both
./verify.sh results/reproduce
\end{lstlisting}

Regenerated paper-facing plots use \artifactpath{paper_legacy}.
\artifactpath{current_corrected} is separate diagnostic output. Fig.~%
\ref{fig:modeling_ref_b200} additionally audits the frozen paper statistic
separately from its legacy rerun. No reported statistic mixes the two paths.

\subsection{Evaluation and expected results}

A successful full run prints \code{workflow: PASS} and
\code{Overall verification: PASS}, then produces 57 validated CSVs, 16 status
markers, and 16 figure groups. The artifact retains accepted-revision stage
IDs: camera-ready Figs.~\ref{fig:modeling_ref_h100},
\ref{fig:modeling_ref_b200}, and \ref{fig:noc_modeling_accuracy} map to
\code{fig08}, \code{fig09}, and \code{fig13}, respectively. Figs.~%
\ref{fig:arch_dse}--\ref{fig:dse-our-ast} map to
\code{fig15}--\code{fig23}. Stages \code{fig10} and \code{fig12} retain the
MI325X and DeepSeek-V3.2 auxiliary validations omitted from the camera-ready
paper.

\begin{itemize}
  \item {\bf Validation (Figs.~\ref{fig:modeling_ref_h100}--%
        \ref{fig:noc_modeling_accuracy}):} H100 weighted MAPE
        $8.4\%$; B200 frozen-paper MAPE $12.92\%$ and legacy-rerun MAPE
        $12.72\%$; and verified NoC closure including the reported
        $108{,}000\times$ endpoint ratio.
  \item {\bf Co-design (Figs.~\ref{fig:arch_dse}--\ref{fig:dse-our-ast} and
        Table~\ref{tab:ablation}):} the workflow regenerates the selected
        winners and sweeps, including the nine-layer bandwidth peak,
        $39.6\%$ decrease by layer 12, and Table~\ref{tab:ablation} speedups
        of $2.79\times$ and $9.47\times$. For Fig.~%
        \ref{fig:e2e_perf_vs_stacked_layers}, panels (a) and (b) combine the
        archived 110-point DSE projection with 44 re-evaluated fixed winners;
        panel (c), the die-area breakdown, is not reproduced.
\end{itemize}

Exact checks and tolerances are in each stage's \artifactpath{summary.csv} and
\artifactpath{docs/result_matrix.md}. The full
$\sim$$2.5\times10^{14}$-point search and the non-distributable
hardware-emulation reference are outside the reproduced scope.

\subsection{Experiment customization}

Custom architecture, energy, NoC, and DSE interfaces and runnable synthetic
examples are documented in \artifactpath{README.md} and
\artifactpath{tests/}. They include caller-owned capacity providers or area
estimators. On failure, inspect the stage log named in
\artifactpath{run_manifest.csv}. For memory pressure, rerun with
\code{--workers 16}.
\endgroup